\documentclass[superscriptaddress,aps, prb,amsmath,amssymb,twocolumn,floatfix]{revtex4}
\usepackage[usenames,dvipsnames]{color}
\usepackage{array}
\usepackage[pdftex]{graphicx}
\usepackage{epsfig}
\usepackage[normalem]{ulem}
\input epsf

\newcommand{\ul}[1]{{\bf #1}}

\newcommand{\boldi}{{\bf i}}
\newcommand{\boldj}{{\bf j}}

\begin{document}
\title{Liquid drops on a surface: using density functional theory to calculate the binding potential and drop profiles and comparing with results from mesoscopic modelling}
\author{Adam P. Hughes}
\email{A.Hughes2@lboro.ac.uk}
\affiliation{Department of Mathematical Sciences, Loughborough University, Loughborough, LE11 3TU, UK}
\author{Uwe Thiele}
\email{u.thiele@uni-muenster.de}
\affiliation{Westf\"alische Wilhelms-Universit\"at M\"unster, Institut f\"{u}r Theorestische Physik, Wilhelm-Klemm-Str.\ 9, 48149 M\"unster, Deutschland}
\affiliation{Center of Nonlinear Science (CeNoS), Westf\"alische Wilhelms Universit\"at M\"unster, Corrensstr. 2, 48149 M\"unster, Germany}
\author{Andrew J. Archer}
\email{A.J.Archer@lboro.ac.uk}
\affiliation{Department of Mathematical Sciences, Loughborough University, Loughborough, LE11 3TU, UK}
\date{\today}

\begin{abstract}
The contribution to the free energy for a film of liquid of thickness $h$ on a solid surface, due to the interactions between the solid-liquid and liquid-gas interfaces is given by the binding potential, $g(h)$. The precise form of $g(h)$ determines whether or not the liquid wets the surface. Note that differentiating $g(h)$ gives the Derjaguin or disjoining pressure. We develop a microscopic density functional theory (DFT) based method for calculating $g(h)$, allowing us to relate the form of $g(h)$ to the nature of the molecular interactions in the system. We present results based on using a simple lattice gas model, to demonstrate the procedure. In order to describe the static and dynamic behaviour of non-uniform liquid films and drops on surfaces, a mesoscopic free energy based on $g(h)$ is often used. We calculate such equilibrium film height profiles and also directly calculate using DFT the corresponding density profiles for liquid drops on surfaces. Comparing quantities such as the contact angle and also the shape of the drops, we find good agreement between the two methods. We also study in detail the effect on $g(h)$ of truncating the range of the dispersion forces, both those between the fluid molecules and those between the fluid and wall. We find that truncating can have a significant effect on $g(h)$ and the associated wetting behaviour of the fluid.
\end{abstract}
\maketitle

\section{Introduction}
The wetting of a substrate by a fluid is an important physical process and understanding such behaviour is crucial in a variety of fields from industrial processes such as lubrication and painting to biological applications such as tear films in the eyes or mucus linings in the lungs. The wetting behaviour of a fluid\cite{degennes85,hansen13}  is determined by the manner in which the atoms or molecules within the fluid interact with each other and with those forming the substrate. Determining the macroscopic fluid properties, wetting behaviour and thermodynamics, starting from an understanding of the (microscopic) molecular interactions is one of the cornerstone problems in liquid state science.\cite{hansen13}

On the macroscopic scale, the wetting behaviour of a fluid in contact with a solid substrate can be characterised by the contact angle that a liquid drop makes with that substrate. Three regimes of wetting behaviour can be identified: complete wetting, partial wetting and non wetting, these three states are defined by contact angles of $\theta=0^\circ$, $0^\circ < \theta < 180^\circ$ and $\theta = 180^\circ$ respectively.\cite{degennes85} Surface tension forces arise from interfaces in the fluid and these can be related to the contact angle by Young's equation\cite{hansen13}
\begin{equation}\label{eq:youngs}
\cos\theta=\frac{\gamma_{wg} - \gamma_{wl}}{\gamma_{lg}},
\end{equation}
where $\gamma_{lg}$, $\gamma_{wl}$ and $\gamma_{wg}$ are the liquid-gas, wall-liquid and wall-gas surface tensions respectively.

The effective {interface} Hamiltonian (IH) model,\cite{dietrich88, schick90, macdowell13b, macdowell14} also referred to as the {interface} free energy model, describes the height profile of a mesoscopic liquid film on a substrate. One can find the equilibrium shape of a droplet by minimising the free energy functional
\begin{equation}\label{eq:fullcurveenergy}
F[h] = \int  \left[ g(h) + \gamma_{lg} \sqrt{1+(\nabla h)^2} \right] \text{d}\ul{x},
\end{equation}
where $h({\bf x})$ is the liquid film thickness at some point $\ul{x}$ on the substrate and $g(h)$ is the binding {potential, which is also referred to as the effective interface potential.\cite{dietrich88, schick90, macdowell02, macdowell13b, macdowell14, henderson05} $g(h)$ is a restricted free energy, i.e.~the free energy subject to the constraint that the thickness of the liquid layer adsorbed on the surfaces is $h$. A good discussion on the subject of restricted free energies can be found in Ref.~\onlinecite{Doi_book}.} The binding potential describes the interaction between two interfaces and is related to the disjoining pressure $\Pi = -\partial g/ \partial h$. In the IH model, the binding potential is often approximated by simple expressions that give the qualitatively correct behaviour. A common example of such an approximation would be an asymptotic expansion which is valid only for larger film thicknesses (cf.\ Sec.\ \ref{sec:binding}). This paper sets out a method to directly calculate the binding potential from a microscopic basis, namely via density functional theory (DFT).\cite{evans79,evans92,hansen13,Lutsko10,Tarazona08,evans10,lowen10} As input, the method takes the interactions between particles in the fluid and also the forces on the fluid particles due to any external fields, for example that due to the wall of a container or a surface on which the liquid is deposited. This calculation yields an expression for the binding potential that is valid for all film thicknesses. The term ``particle'' is used here generically to refer to the atoms/molecules/colloids in the fluid, depending on the exact system under study. Note that it is also possible to calculate $g(h)$ from computer simulations -- see Refs.\,\onlinecite{macdowell05, macdowell06, herring10, macdowell11, rane11, gregorio12, tretyakov13, benet14}. 

\begin{figure}
\vspace{-2em}
\includegraphics[width=1.05\columnwidth]{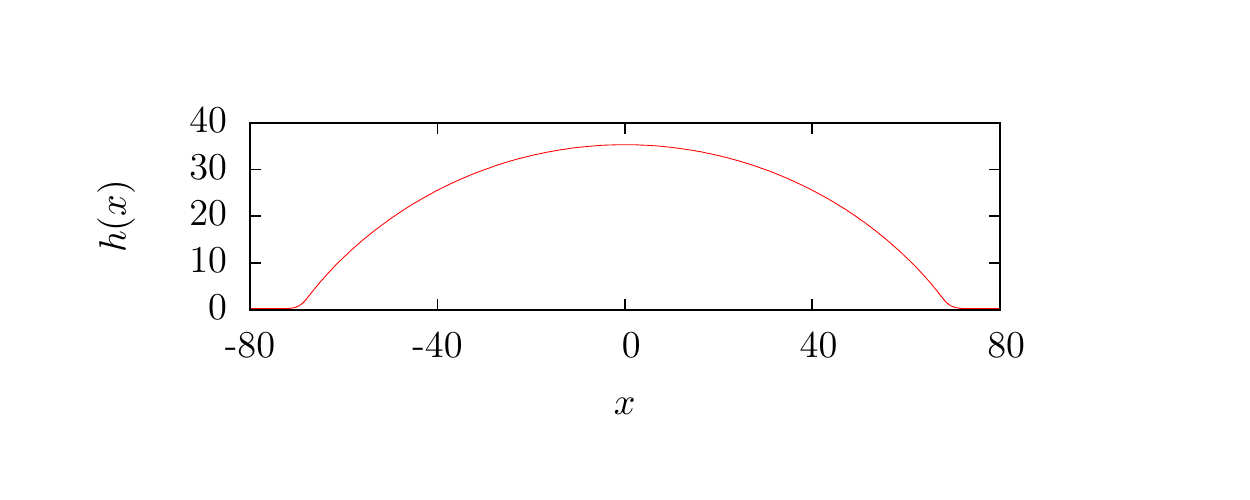}

\vspace{-3em}
\includegraphics[width=1.05\columnwidth]{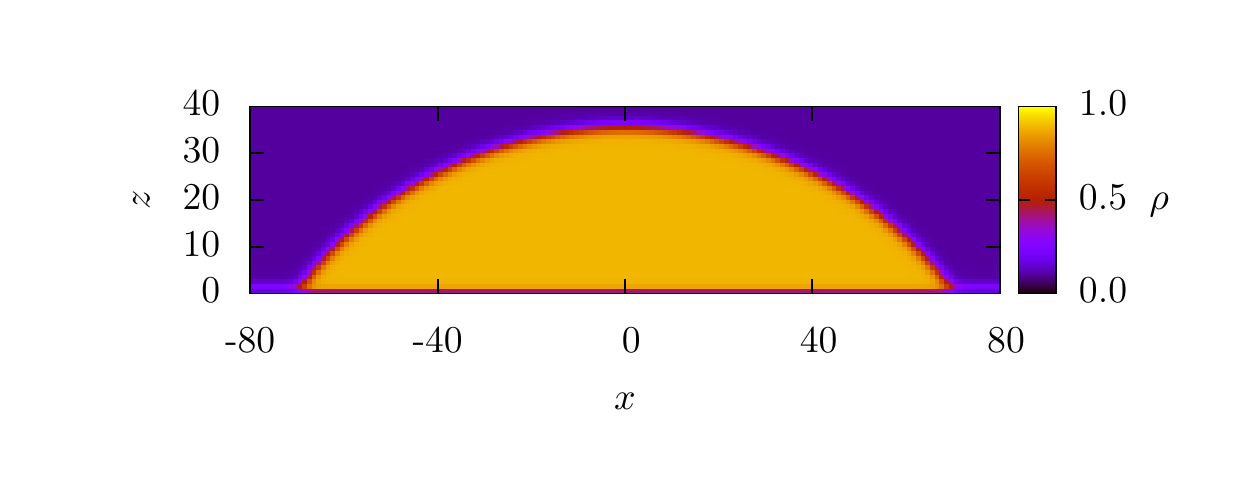}

\vspace{-2em}
\caption{Two different descriptions of a liquid drop on a surface: (a) a height profile calculated via Eq.\ \eqref{eq:filmHeight} from a mesoscopic free energy (cf.\ Eq~\eqref{eq:fullcurveenergy}) and (b) a density profile, which gives the fluid number density at a distance $z$ above the surface. These are both for a fluid with $\beta \epsilon = 0.9$ and $\beta \epsilon_w = 0.6$ (see Sec.\ \ref{sec:lattice} for further details).}
\label{fig:profs}
\end{figure}

The IH model is particularly useful due to its application in dynamical studies. By making the assumption of small surface gradients and contact angles, Eq.\,\eqref{eq:fullcurveenergy} reduces to
\begin{equation}\label{eq:thinEnergy}
F[h] = \int \left[g(h) + \frac{\gamma_{lg}}{2} (\nabla h)^2 \right] \text{d}\ul{x}
\end{equation}
where we have omitted a constant contribution. Equation~\eqref{eq:thinEnergy} can then be employed in the thin film (or long-wave) evolution equation,\cite{oron97,thiele07} which describes the time evolution of a thin film of liquid on a flat solid substrate. In gradient dynamics form it is written\cite{mitlin93,thiele10}
\begin{equation}\label{eq:thinFilm}
\frac{\partial h}{\partial t} = \nabla \cdot \left[ Q(h) \nabla \frac{\delta F[h]}{\delta h} \right],
\end{equation}
where $Q(h)$ is a mobility factor that depends on the film thickness $h(\ul{x},t)$. Equation~\eqref{eq:thinFilm} may be derived by making a long wave approximation in the governing Navier-Stokes equation.\cite{safran94, thiele07} There are many applications of this equation to model different situations. Steady state solutions, where $\partial h/\partial t=0$, such as drop profiles, are found by minimising the free energy, Eq.\,\eqref{eq:thinEnergy}, with respect to the film height profile, subject to a volume constraint. More specifically, it amounts to solving
\begin{equation}
\frac{\delta F}{\delta h} = \alpha,
\end{equation}
where $\alpha$ is a Lagrange multiplier stemming from the constraint
\begin{equation}
\int \text{d} \ul{x} \  h(\ul{x}) = V_0,
\end{equation}
where $V_0$ is a specified drop volume. A typical drop profile resulting from such a calculation can be seen in Fig.\,\ref{fig:profs}(a). Note the very thin non-zero height  `precursor' film that is present to the left and right of the drop.

A fully microscopic description has statistical mechanics as its basis. Statistical mechanics calculates an average over an ensemble of all possible states of the system, i.e.\ it averages over all possible configurations of the particles. This average leads to determining the fluid one body density profile $\rho(\ul{r})$, which represents the likelihood of finding a particle at a given point $\ul{r}$ in the system.\cite{hansen13} This statistical mechanical point of view is the basis for DFT. From DFT, the grand potential, $\Omega$, of the system is calculated and the equilibrium density profile, $\rho(\ul{r})$, which minimises $\Omega$, can be found. A typical equilibrium density profile of a liquid drop on a solid substrate is displayed in Fig.\,\ref{fig:profs}(b). Other examples of drop profiles calculated using DFT can be found in Refs.\,\onlinecite{berim08, ruckenstein10, pereira11, nold11, nold14}. Note that by identifying the surface of the liquid as the surface where the density equals a specified value, $\rho_{int}$, where $\rho_l>\rho_{int}>\rho_g$ and where $\rho_l$ and $\rho_g$ are the coexisting liquid and gas densities of the fluid, the description can be further reduced to obtain a film height profile very similar to that displayed in Fig.\,\ref{fig:profs}(a). One possible choice is to choose $\rho_{int}=(\rho_l+\rho_g)/2$. Alternatively, by integrating in the $z$ direction over such a density profile we can obtain a drop height profile. Here, we define the height of the liquid film on a substrate as the adsorption divided by the liquid-gas bulk density difference
\begin{equation}\label{eq:filmHeight}
h(x,y)=\frac{\int_0^\infty \left[\rho(x,y,z)-\rho_g \right] dz}{\rho_l - \rho_g}.
\end{equation}
It is worth noting that using DFT it is just as easy to compute the profile for a liquid droplet that makes a contact angle with the substrate that is greater than $90^\circ$, than one with $\theta < 90^\circ$. It is significantly more difficult to find drop profiles for $\theta > 90^\circ$ in the mesoscopic approach because for these contact angles, Eqs.\ \eqref{eq:fullcurveenergy} and \eqref{eq:thinEnergy} cannot be employed.

The remainder of this paper is set out as follows: in Sec.~\ref{sec:binding}, the binding potential and the procedure for calculating it are discussed. A simple DFT, for the lattice-gas model, is presented in Sec.\ \ref{sec:lattice} that is used to demonstrate the procedure. The dependence of the fluid behaviour on the particle interactions is discussed in Sec.~\ref{sec:ranges}. In particular, it is shown that truncating the range of the dispersion interactions between the particles has a profound effect on the binding potential and interfacial phase behaviour. The method of fitting a function to the calculated data is given in Sec.~\ref{sec:fitting}, followed by the results of passing the binding potential from the lattice-gas model to the thin film IH model in Sec.~\ref{sec:results}. Finally, conclusions are drawn in Sec.~\ref{sec:conclusion}.

\section{The Binding Potential}\label{sec:binding}
\begin{figure}
\begin{center}
\includegraphics{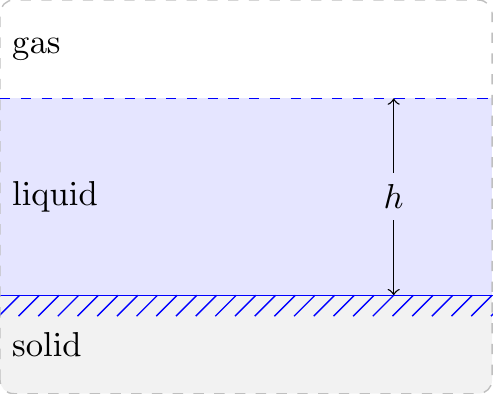}
\caption{A schematic of the system: a film of liquid of thickness $h$ separating a semi-infinite volume of gas from a solid surface.}
\label{fig:system}
\end{center}
\end{figure}

For any fluid, coexistence between the liquid and gas phases occurs when the temperature $T$, chemical potential $\mu$ and pressure $p$ of the two phases are equal. It then follows, for a given volume $V$, that the bulk grand free energy, $\Omega=-pV$, of either phase occupying the same volume is also equal. For a system where the liquid and gas phases of a fluid exist together at the point of liquid-gas coexistence, any excess, over bulk, contributions to the free energy of the system must stem solely from the interface that forms between the two phases. This excess grand potential per unit area of the interface defines the surface tension between those two phases. In this case, it is the liquid-gas surface tension $\gamma_{lg}$. Now consider a system with chemical potential $\mu=\mu_{coex}$, the value at coexistence, where a film of liquid separates the bulk gas from a solid surface (cf.\ Fig.\,\ref{fig:system}). The excess free energy now consists of the sum of the two interfacial tensions and also the interaction between the two interfaces. The excess grand potential in such a system is given by $\Omega_{ex}(h)\equiv\Omega+pV=\omega_{ex}(h)A$, where $A$ is the area of the interface and
\begin{equation}\label{eq:freeEnergy}
\omega_{ex}(h)=\gamma_{wl}+\gamma_{lg}+g(h).
\end{equation}
$h$ is the liquid film thickness, $\gamma_{wl}$ and $\gamma_{lg}$ are the wall-liquid and liquid-gas surface tensions respectively.\cite{marchand11} The final term $g(h)$ is the binding potential, which gives the contribution to the free energy from the interaction between the two interfaces. This has the property that as $h \to \infty$, $g(h) \to 0$. The absolute minimum of the grand potential defines the equilibrium film thickness $h$.

In the case of liquid droplets surrounded by their vapour on a solid substrate, the absolute minimum of the binding potential is directly related to the equilibrium contact angle of the drop,\cite{churaev95a,rauscher08}
\begin{equation}\label{eq:bindCont}
\theta=\cos^{-1}\left(1+\frac{g(h_0)}{\gamma_{lg}}\right),
\end{equation}
where $g(h_0)$ is the value at the minimum of the binding potential. This corresponds directly to Young's equation, Eq.\,\eqref{eq:youngs}; the absolute minimum of the binding potential corresponds to the equilibrium state of the system and gives the equilibrium excess grand potential, i.e.\ the wall-gas surface tension:
\begin{equation}
\omega(h_0) = \gamma_{wl}+ \gamma_{lg} + g(h_0) =  \gamma_{wg}.
\end{equation}
Replacing $\gamma_{wg}$ in Eq.\,\eqref{eq:youngs} with this expression leads to Eq.\,\eqref{eq:bindCont}.

Equation~\eqref{eq:freeEnergy} is given above as a function of the film height $h$. From a microscopic viewpoint it is often more convenient to use the adsorption of the fluid as the order parameter characterising the fluid at the interface, instead of the film thickness. The total adsorption is readily calculated from the fluid density profile as
\begin{equation}\label{eq:adsorption}
\Gamma=\frac{1}{A} \int \text{d}\ul{r} (\rho(\ul{r}) - \rho_b).
\end{equation}
$\rho_b$ is the bulk fluid density, which in the cases considered here is the gas density $\rho_g$. We may also define a local adsorption [c.f.~Eq.~\eqref{eq:filmHeight}] as follows:
\begin{equation}\label{eq:filmAds}
\Gamma(x,y)=\int_0^\infty \left[\rho(x,y,z)-\rho_g \right] dz,
\end{equation}
so that when the definition in Eq.~\eqref{eq:filmHeight} for the film height $h$ is used, we have $\Gamma = h(\rho_l- \rho_g)$. For other definitions of the film height $h$, then 
\begin{equation}\label{eq:adsorption_approx}
\Gamma \approx h(\rho_l- \rho_g).
\end{equation}
The dependence on how precisely $h$ is defined becomes negligible, for large $h$. Thus, Eq.~\eqref{eq:freeEnergy} and also the binding potential $g$ may be given as a function of the adsorption. Note, however, that negative adsorptions are possible, e.g.\ at a purely repulsive wall, and in such a situation describing the liquid film at the wall via a film thickness $h$ becomes meaningless since the quantity $h$ defined in Eq.~(\ref{eq:filmHeight}) then becomes negative.

For a system with chemical potential $\mu=\mu_{coex}+\delta\mu$, i.e.\ off-coexistence, then there is an additional contribution $\Gamma\delta\mu$, that must be added to the right hand side of Eq.~\eqref{eq:freeEnergy}.\cite{evans89, schick90} Together with Eq.\ \eqref{eq:adsorption_approx}, this gives
\begin{equation}\label{eq:freeEnergy_with_dmu}
\omega_{ex}(h)\approx\gamma_{wl}+\gamma_{lg}+(\rho_l-\rho_g)h\delta\mu+g(h).
\end{equation}

Often, only asymptotic forms of binding potentials are used which are strictly valid only in the limit of a large film thickness.\cite{schick90} There are two main asymptotic forms of the binding potential that are considered, the choice of which depends on the assumed particle interactions and the range of those interactions. For van der Waals' (dispersion) interactions, the following asymptotic form is appropriate:\cite{dietrich88,ibagon13,ibagon14,stewart05b}
\begin{equation}\label{eq:longAsym}
g(h)=\frac{a}{h^2} + \frac{b}{h^3} + \cdots,
\end{equation}
The equivalent disjoining pressure is\cite{thiele07,oron97}
\begin{equation}
\Pi(h)= \frac{2a}{h^3}+ \frac{3b}{h^4}+ \cdots.
\end{equation}
Under certain approximations the asymptotic behaviour shown in Eq.\,\eqref{eq:longAsym} can be calculated analytically including the values of the coefficients $a$ and $b$, and how they depend on the temperature.

With only short ranged interactions between particles, the binding potential can be expressed asymptotically as\cite{fernandez11,henderson05,brezin83,macdowell02,dietrich88}
\begin{equation}\label{eq:shortAsym}
g(h)=a \exp(-h/ \xi) + b \exp(-2h/\xi) + \cdots.
\end{equation}
The length $\xi$ is the bulk correlation length in the liquid phase at the interface. These asymptotic forms, truncated after a few terms, are used frequently throughout the literature (independently or as combinations of the two forms\cite{degennes85, thiele07, evans09, frastia12}) even though they are only strictly valid for thick liquid films and can not describe the binding potential as $h\to 0$. To describe the small $h$ behaviour, the full form of $g(h)$ is required. A fully microscopic theory, that describes the fluid structure at the wall, is required to obtain this.
\begin{figure}
\begin{center}
\includegraphics[width=0.48\textwidth]{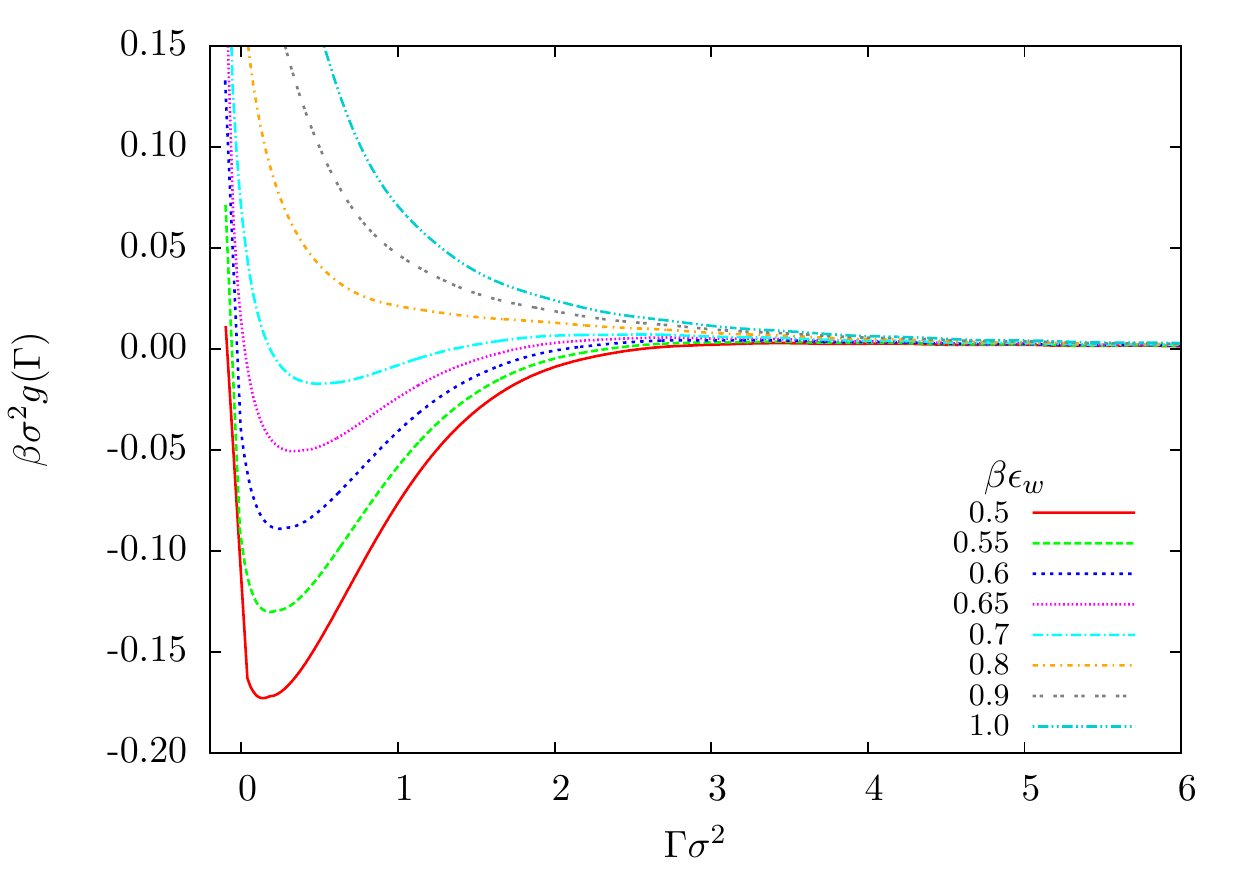}
\caption{A series of binding potentials for a fluid with fixed inverse temperature $\beta \epsilon=0.9$ against a solid substrate with varying attraction strength $\beta \epsilon_w$. A change in wetting behaviour, from non-wetting to wetting occurs, as indicated by the change in the position of the minimum in $g(\Gamma)$ from a low finite adsorption to a large $(\Gamma \to \infty)$ adsorption, as $\beta \epsilon_w$ increases.}
\label{fig:bindingPot}
\end{center}
\end{figure}

Fig.~\ref{fig:bindingPot} shows binding potentials for various values of a parameter $\epsilon_w$, which determines whether or not the fluid wets the wall. These binding potentials are calculated using the microscopic DFT-based approach that is introduced below (cf. Sec.\ \ref{sec:lattice}). The parameter $\epsilon_w$ characterises the interaction strength between the wall and the fluid particles. The full form of the long range (algebraic) dispersion interactions between the particles are included and so for large $\Gamma$, $g(\Gamma)$ has the asymptotic form given in Eq.\,\eqref{eq:longAsym}. In Fig.~\ref{fig:bindingPot} we see that for small values of $\epsilon_w$ (weakly attracting, solvophobic wall) the global minimum in $g$ is at a small value of $\Gamma$, i.e.\ the liquid does not wet the wall. As $\epsilon_w$ is increased, there is a first order wetting transition when $\beta\epsilon_w\approx0.74$ and for $\beta\epsilon_w>0.74$ the global minimum in $g(\Gamma)$ is at $\Gamma\to\infty$, i.e.\ there is a macroscopically thick film of liquid on the wall at coexistence $\mu=\mu_{coex}$.

DFT gives a route by which the free energy may be calculated, taking into account the microscopic structure of the fluid at the wall and the interactions of the particles within it. The equilibrium state of the system is found by minimising the grand potential functional:\cite{evans79,evans92,hansen13,Lutsko10,Tarazona08,evans10,lowen10}
\begin{equation}
\Omega[\rho({\bf r})] = {\cal F}[\rho({\bf r})] + \int \text{d}{\bf r} V_{ext}({\bf r})\rho({\bf r}) - \mu \int \text{d}{\bf r} \rho({\bf r}),
\end{equation}
where ${\cal F}$ is the intrinsic Helmholtz free energy functional, $\mu$ is the chemical potential and $V_{ext}$ is the external potential. The equilibrium density profile, $\rho(\ul{r})$, is that which satisfies the Euler-Lagrange equation
\begin{equation}
\frac{\delta \Omega[\rho]}{\delta \rho({\bf r})}=0.
\label{eq:EL_eq}
\end{equation}
Once an approximation for ${\cal F}$ is specified, this equation is solved numerically via a scheme in which an initial density profile is supplied and then iterated until a specified convergence criterion is reached.\cite{hughes13,roth10}

In order to calculate the binding potential as a function of the adsorption using DFT, it is required to evaluate the free energy for any specified $\Gamma$, i.e.\ in addition to the equilibrium profile we require other non-equilibrium profiles for a range of values of the adsorption $\Gamma$. By using the procedure developed and justified in Ref.~\onlinecite{archer11} (see also Ref.~\onlinecite{archer13}), the excess density $(\rho({\bf r}) - \rho_b)$ is normalised at each iterative step, which is equivalent to including a (fictitious) additional effective external potential that stabilises a wetting film of the desired thickness with the specified value of $\Gamma$. Using this normalisation procedure, the excess grand potential can be obtained for a range of values of $\Gamma$. The binding potential is then given via Eq.~\eqref{eq:freeEnergy} by subtracting the values of the solid-liquid and liquid-gas surface tensions.

\section{A Microscopic Model}\label{sec:lattice}
The method outlined above for calculating the binding potential is valid for any DFT model. To illustrate the procedure, a simple approximate DFT for a lattice-gas, is used. The model is only briefly described here: a full description and derivation can be found in Ref.~\onlinecite{hughes13}. The system is discretised by a cubic lattice and the fluid particles, assumed to all be identical and spherical,  occupy only one cell each on the lattice. The diameter of each particle, and the width of each cell, is $\sigma$ and there are $M=M_i M_j M_k$ lattice cells and $N<M$ fluid particles in the system. A point in the lattice is denoted ${\bf i}=(i,j,k)$ and  $M_i$, $M_j$, and $M_k$ are the number of cells in the $i$, $j$ and $k$ directions, respectively. Note $i$, $j$ and $k$ are integers. Any configuration of particles in the system can then be described by the set of occupation numbers, $\{ n_\boldi\}$, where $n_\boldi=1$ if there is a particle in cell $\boldi$ and $n_\boldi=0$ otherwise. The average of an ensemble of all such systems may be taken and then the average occupation number of each cell is described by the density $\rho_\boldi=\langle n_\boldi \rangle$. It may be assumed that the equilibrium density profile only varies in one direction, or, as for the calculations below, it is assumed that the density profile is invariant in the third dimension, so only a two-dimensional (2D) slice of the system needs to be studied. Here, the $z$-axis direction, indexed by $k$, is assumed to be invariant. The energy of the full three-dimensional system is given by the Hamiltonian
\begin{equation}\label{eq:hamiltonian}
E= - \frac12 \sum_{\boldi=1}^M \sum_{\boldj \neq \boldi} \tilde{\epsilon}_{\boldi,\boldj} n_\boldi n_\boldj + \sum_{\boldi=1}^M \tilde{V}_\boldi n_\boldi,
\end{equation}
where $\tilde{V}_\boldi$ is the external potential and $\tilde{\epsilon}_{\boldi,\boldj}$ is a Lennard-Jones-like pair potential between particles at lattice sites $\boldi$ and $\boldj$: 
\begin{equation}
\tilde{\epsilon}_{\boldi,\boldj} = v(r_{\boldi,\boldj}) =
\begin{cases}
- \epsilon / {r_{\boldi,\boldj}}^6 & \text{ for } r_{\boldi,\boldj} \geq \sigma, \\
\infty & \text{ for } r_{\boldi,\boldj} < \sigma,
\end{cases}\label{eq:pair_pot}
 \end{equation}
where $r_{\boldi,\boldj}$ is the distance between a pair of particles located at lattice sites $\boldi$ and $\boldj$. If one assumes that the system is invariant along the direction indexed by $k$, then a mean-field approximation for the internal energy $U$ is the following 2D sum:
\begin{equation}
U = \langle E\rangle = -\frac12 \sum_{\boldi_0=1}^{M_{2d}} \sum_{\boldj_0 \neq \boldi_0} \epsilon_{\boldi_0,\boldj_0} \rho_{\boldi_0} \rho_{\boldj_0} + \sum_{\boldi_0=1}^{M_{2d}} V_{\boldi_0} \rho_{\boldi_0},
\end{equation}
where $\sum_{\boldi_0=1}^{M_{2d}}$ sums over the $M_{2d}=M_iM_j$ sites in the 2D lattice plane where $k=0$ and $\boldi_0=(i,j,0)$. The third (invariant) dimension is now accounted for in the interaction weights $\epsilon_{\boldi,\boldj}$ and $V_\boldi$. The sums are written explicitly in Eqs.\,\eqref{eq:external} to \eqref{eq:truncWeights} {in Appendix \ref{app:appB}, these equations also give the weights when the range of particle interactions is truncated.} The grand potential of this system can be approximated as follows:\cite{hughes13}
\begin{align}\label{eq:latGas}
\Omega(\{\rho_{\boldi_0}\}) & =  U-TS-\mu N, \nonumber \\
	& = k_BT \sum_{\boldi_0=1}^{M_{2d}} \left[ \rho_{\boldi_0} \ln(\rho_{\boldi_0})+(1- \rho_{\boldi_0})\ln(1- \rho_{\boldi_0})\right] \nonumber \\
	& \qquad- \frac12 \sum_{\boldi_0=1}^{M_{2d}} \sum_{{\boldj_0} \neq {\boldi_0}} \epsilon_{{\boldi_0},{\boldj_0}} \rho_{\boldi_0} \rho_{\boldj_0} + \sum_{\boldi_0=1}^{M_{2d}} \rho_{\boldi_0}(V_{\boldi_0} - \mu),
\end{align}
where $k_B$ is Boltzmann's constant and $S$ is the entropy\footnote{Note that the contribution to the lattice gas free energy density $\beta f=\rho\ln\rho-(1-\rho)\ln(1-\rho)$ in Eq.~\eqref{eq:latGas} contains both the ideal gas $\beta f_{id}=\rho\ln\rho-\rho$ and an excess contribution $\beta f_{ex}=\rho+(1-\rho)\ln(1-\rho)$; i.e.\ $f=f_{id}+f_{ex}$. The origin of this form arises from the hole-particle $n_\boldi\leftrightarrow1-n_\boldi$ symmetry in the lattice model (see e.g.\ Ref.\,\onlinecite{hughes13}). The second term, $f_{ex}$, stems from the particles having a hard-core interaction, preventing multiple occupancy of a cell. It is instructive to compare this with the free energy of a binary mixture. The ideal-gas contribution to the free energy of a mixture is $\beta f=\rho_1\ln\rho_1-\rho_1+\rho_2\ln\rho_2-\rho_2$, where $\rho_1$ is the density of species 1 and $\rho_2$ the density of species 2. Re-writing this in terms of the total density $\rho=\rho_1+\rho_2$ and concentration $x=\rho_1/\rho$, we obtain $\beta f=\rho\ln\rho-\rho+\rho[x\ln x+(1-x)\ln(1-x)]$. The final term is the ideal-gas contribution to the entropy of mixing and it is interesting to note the similarity in form to the lattice-gas free energy. For an incompressible binary mixture $\rho$ is a constant and so the first two terms are often neglected. As a consequence of this, the free energy of an incompressible binary fluid can have a very similar mathematical form to the lattice-gas free energy considered here.}. Included implicitly in the above equation is the particle diameter $\sigma=1$. 

The set of lattice densities that describes the system at equilibrium is found by solving [c.f.\ Eq.\ \eqref{eq:EL_eq}]:
\begin{equation}\label{eq:eqlbrmCon}
\frac{\partial \Omega}{\partial \rho_\boldi}=0,
\end{equation}
for every lattice site $\boldi$. From Eqs.\,\eqref{eq:latGas} and \eqref{eq:eqlbrmCon} we obtain
\begin{equation}\label{eq:difLatGas}
k_BT\ \ln\left(\frac{\rho_\boldi}{1-\rho_\boldi}\right) - \sum_{\boldj=1}^{M_iM_j} \epsilon_{\boldi,\boldj} \rho_\boldj + V_\boldi-\mu = 0.
\end{equation}
Solving this coupled set of equations \eqref{eq:difLatGas} gives the equilibrium fluid profile $\{\rho_\boldi\}$. However, in order to obtain the non-equilibrium profile with specified adsorption $\Gamma_d$, the iterative method developed in Ref.\,\onlinecite{archer11} must be used. This is equivalent to solving the set of coupled equations [c.f.\ Eq.\,\eqref{eq:difLatGas}]
\begin{equation}
k_BT\ \ln\left(\frac{\rho_\boldi}{1-\rho_\boldi}\right) - \sum_{\boldj=1}^{M_iM_j} \epsilon_{\boldi,\boldj} \rho_\boldj + V_\boldi + V_\boldi^\text{eff} -\mu = 0,
\end{equation}
where $V_\boldi^\text{eff}$ is the fictitious additional potential mentioned at the end of Sec.\ \ref{sec:binding}. It has the property that $V_\boldi^\text{eff} \to 0$ as $i \to \infty$, far from the wall. This additional potential stabilises a film of liquid with the specified adsorption against the wall. Note that $V_\boldi^\text{eff}$ is a-priori unknown, it is calculated on-the-fly self consistently as part of the minimisation algorithm.\cite{archer11} This is done by re-normalising the density profile during every iteration of the algorithm by replacing the value of the density with
\begin{equation}\label{eq:renorm}
\rho_\boldi^\text{new} = \frac{\Gamma_d}{\Gamma_\text{old}} (\rho_\boldi^\text{old} - \rho_b) + \rho_b,
\end{equation}
where $\Gamma_d$ is the desired adsorption and $\Gamma_\text{old}$ is the adsorption corresponding to the density profile $\{\rho_\boldi^\text{old}\}$, obtained from iterating Eq.\,\eqref{eq:difLatGas}. More details about this algorithm and its properties can be found in Ref.\ \onlinecite{archer11}.

{Note that this procedure does not require the bulk phase to be at coexistence, although in all the results presented here it is at coexistence. As the bulk fluid state point $(\mu,T)$ is varied, the form of the restricted free energy $g(\Gamma)$ also changes as well as the bulk densities and, in consequence, the interface tensions. Note also that one can vary the adsorption $\Gamma$ by the standard method of varying the value of the chemical potential $\mu$. Calculations (not presented here) show that the main difference between results from our method for varying the film thickness via a fictitious external potential, with results from varying it by changing $\mu$ are to be seen when the adsorption is small. This is particularly so when the fluid wets the wall, because in this case, to obtain a small adsorbed film height by varying the chemical potential requires a large shift in the value away from that at coexistence.} 

\begin{figure}
\begin{center}
\includegraphics[width=0.48\textwidth]{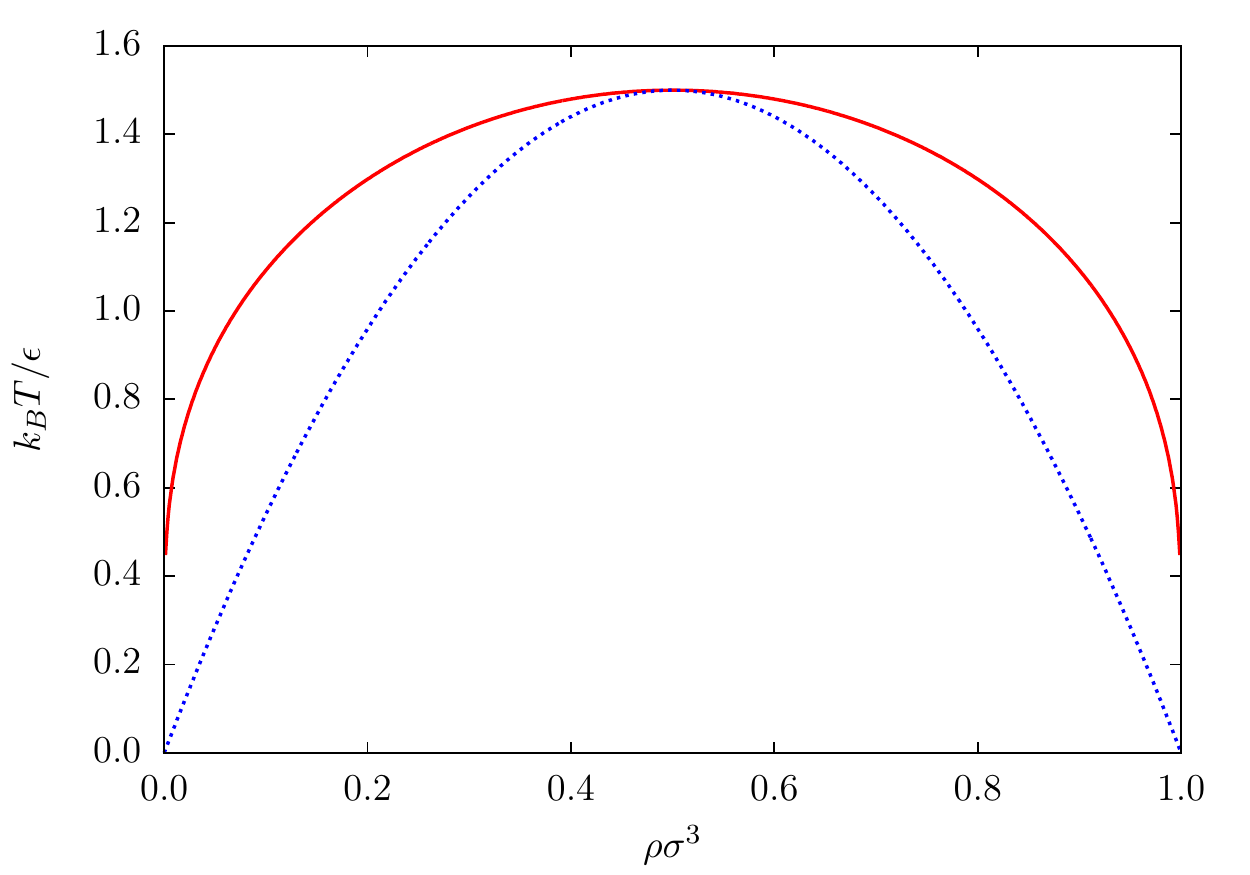}
\caption{The phase diagram for the lattice fluid in the temperature $k_BT/\epsilon$ versus density $\rho\sigma^3$ plane. The solid (red) line is the binodal curve and the dashed (blue) line is the spinodal curve.}
\label{fig:phase}
\end{center}
\end{figure}

The bulk fluid phase diagram is displayed in Fig.\,\ref{fig:phase}. For details about how this is calculated see e.g.\ Refs.\, \onlinecite{hughes13,robbins11}. The binodal and spinodal are both displayed. The binodal gives the densities of the coexisting gas and liquid states. Within the spinodal curve, the uniform fluid is unstable and spontaneous demixing occurs. The bulk critical point is at $\rho\sigma^3=0.5$ and temperature $k_BT/\epsilon=1.5$.

\begin{figure*}
\includegraphics[width=0.66\columnwidth]{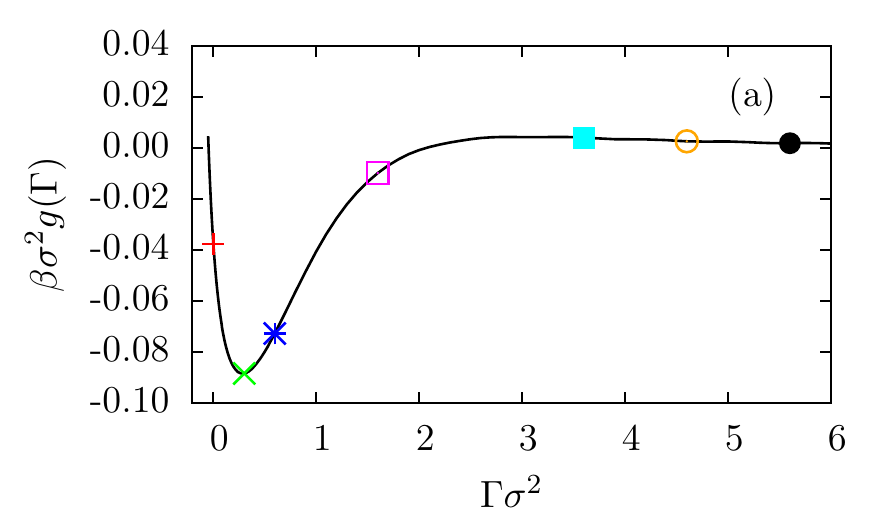}
\includegraphics[width=0.66\columnwidth]{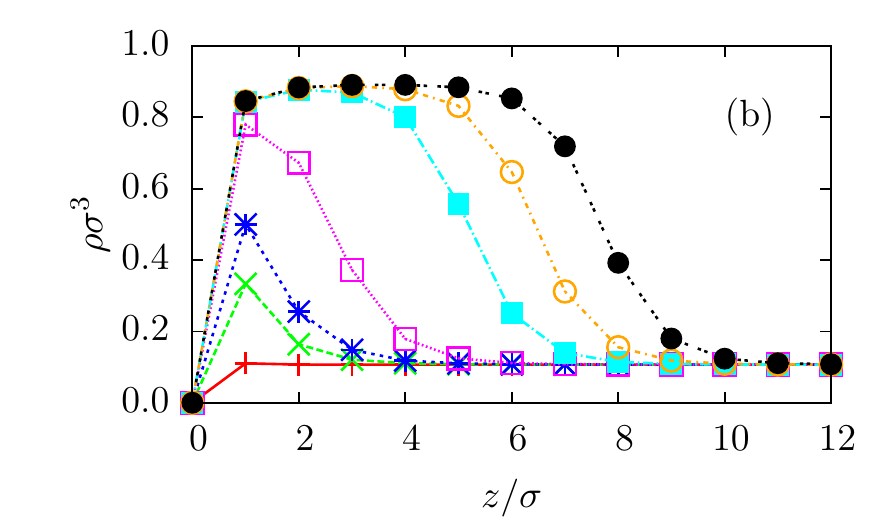}
\includegraphics[width=0.66\columnwidth]{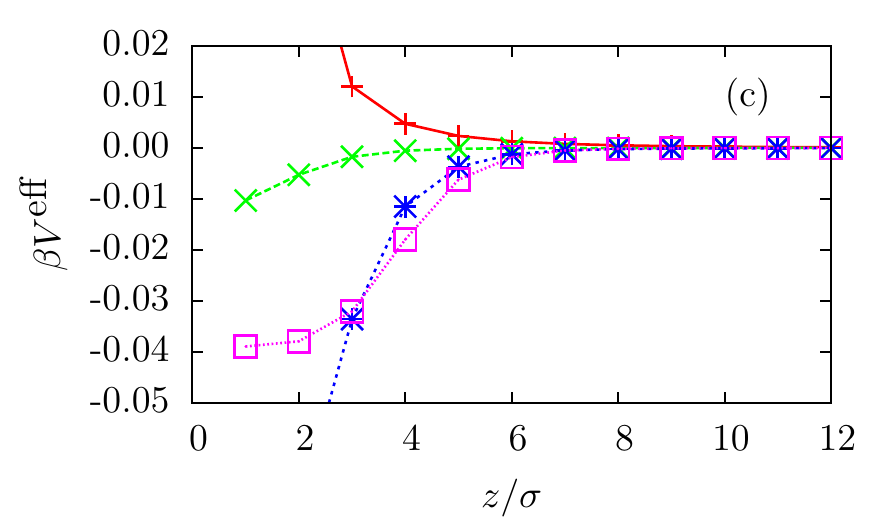}

\includegraphics[width=0.66\columnwidth]{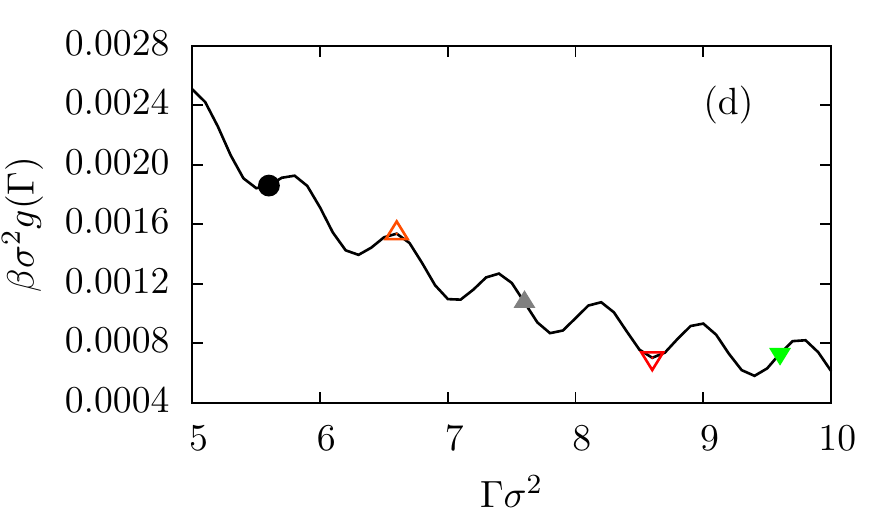}
\includegraphics[width=0.66\columnwidth]{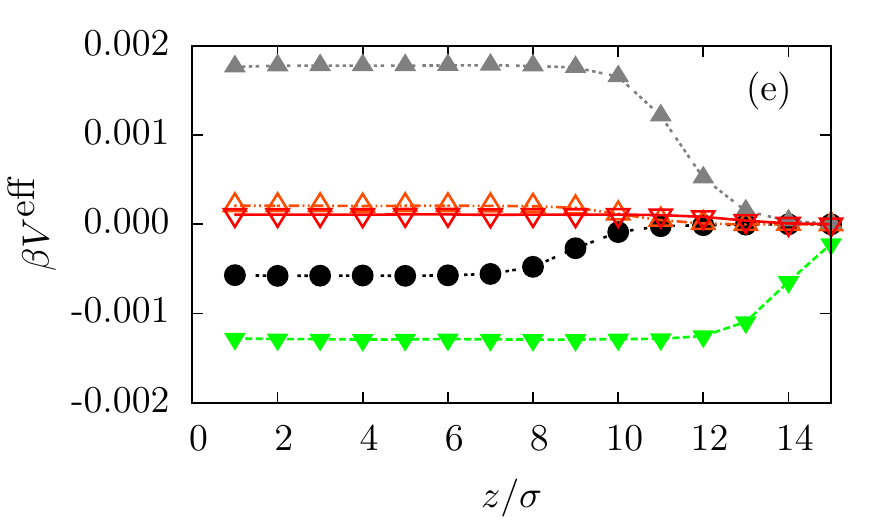}
\caption{In (a) we display the binding potential $g(\Gamma)$ for a fluid with bulk gas density $\rho \sigma^3=0.107$ and temperature $\beta \epsilon=0.9$ against a planar wall with attraction strength $\beta \epsilon_w=0.6$. In (d) we display a magnification of tail of $g(\Gamma)$, for larger values of the adsorption $\Gamma$. Marked on $g(\Gamma)$ are points that correspond to the density profiles displayed in (b), and the corresponding fictitious potentials $V_\boldi^\text{eff}$, which are displayed in (c) and (e). The marked points have adsorption values: $\Gamma \sigma^2 = 0.004$, 0.304, 0.6, 1.6, 3.6, 4.6, 5.6, 6.6, 7.6, 8.6, 9.6 (note that the density profiles for the final four values are not displayed in (a) and (b), for clarity). The fictitious potential is that which must be applied to stabilise a film of liquid with the given adsorption in an open (grand-canonical) system. We observe a clear relation between the gradient of $g(\Gamma)$ and the magnitude and sign of the fictitious external potential.}
\label{fig:prof1d}
\end{figure*}

DFT is a statistical mechanical theory - i.e.\ in principle it should give the ensemble average density profile of the fluid. A statistical description of a fluid confined in an external potential should yield an (ensemble average) density profile with the same symmetry as that potential.\cite{evans79,malijevsky13} Thus, for a planar wall, the equilibrium density profiles only vary with the distance from the wall. Fig.\,\ref{fig:prof1d} shows typical examples of such profiles for the lattice-gas model and of the corresponding points on the binding potential that they represent. This series of density profiles range from small (including negative) values of the adsorption to large values, where the profiles indicate that there is a thick film of liquid at the wall. 

The corresponding fictitious potentials are also displayed in Fig.\,\ref{fig:prof1d}. At the minimum in $g(\Gamma)$, the fictitious potential $V_\boldi^\text{eff}$ is, of course, zero, because this corresponds to the equilibrium state. Moving away to either side of this minimum, $V_\boldi^\text{eff}$ increases rapidly and the largest magnitude potentials are observed for low adsorptions $\Gamma$, for values of $\Gamma$ where the gradient in $g(\Gamma)$ is largest. In Fig.\,\ref{fig:prof1d} the density profile (and associated potential $V_\boldi^\text{eff}$) corresponding to $\Gamma \sigma^2 = 0.304$ is very close to at equilibrium and so $V_\boldi^\text{eff}$ is very small everywhere. In contrast, the fictitious potential for $\Gamma \sigma^2=0.004$ is much larger, as the gradient of $g(\Gamma)$ at this point is also large. For larger adsorption values, in the tail of the binding potential, $V_\boldi^\text{eff}$ can be either weakly attractive or weakly repulsive. This is due to the oscillations in $g(\Gamma)$ stemming from the fact that we are dealing with a lattice model. In a continuum DFT model these oscillations decrease in amplitude as $\Gamma$ increases or are entirely absent, depending on the fluid state point. In the subsection below, we discuss this issue further.

Note that even when the typical microstates of the system consist of liquid drops on the surface, after performing a statistical average over all states, the resultant density profile should be invariant in the direction parallel to the substrate.\cite{malijevsky13} In order to study liquid drops or liquid ridges (invariant in one direction along the surface) one must break the translational symmetry and impose that the centre of mass be located at a particular point or line on the surface. By constraining the centre of mass, 2D drop profiles (i.e.\ liquid ridges) can also be found for the same system even though the external potential only varies in one direction. 

\begin{figure}
\includegraphics[width=\columnwidth]{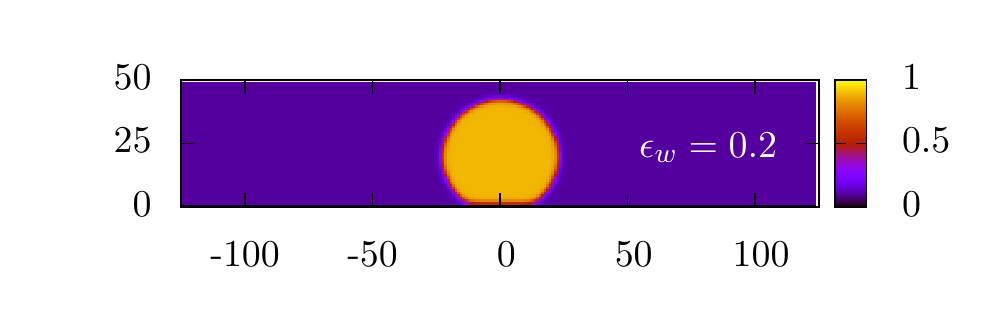}
\includegraphics[width=\columnwidth]{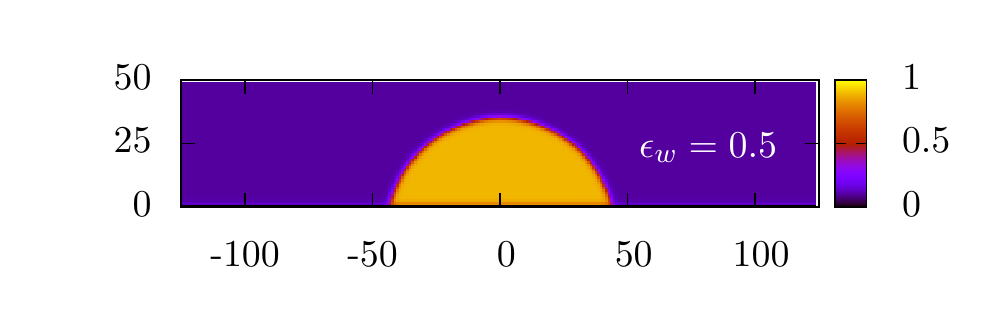}
\includegraphics[width=\columnwidth]{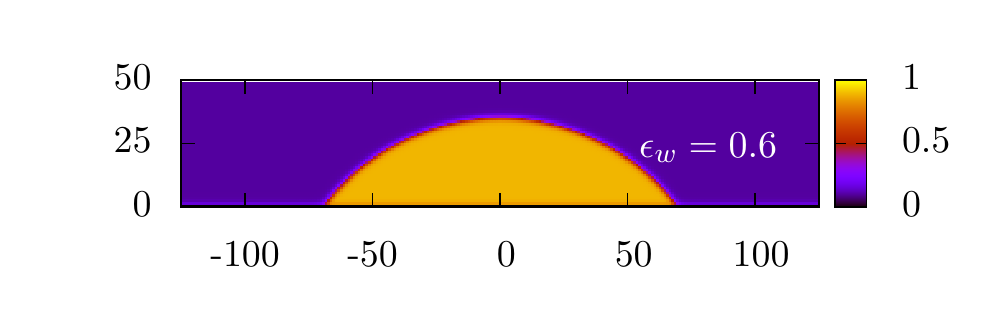}
\includegraphics[width=\columnwidth]{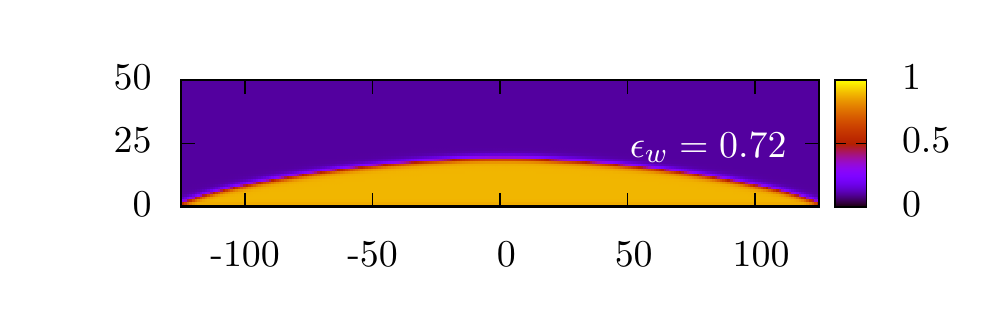}
\includegraphics[width=\columnwidth]{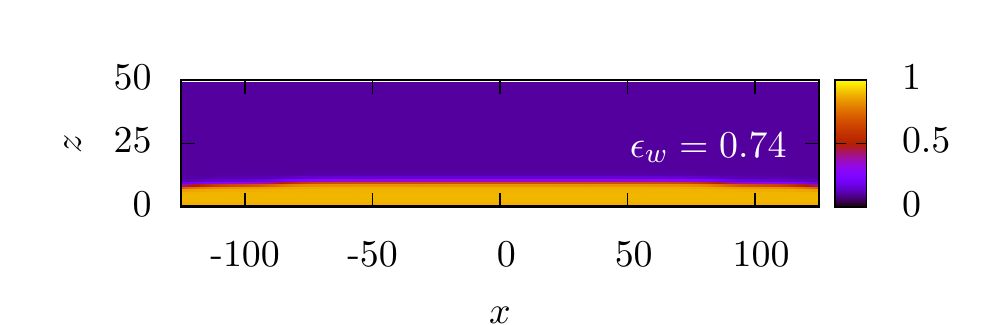}
\caption{A series of 2D density profiles for a drop of the lattice-gas fluid with $\beta \epsilon=0.9$, deposited on various solid substrates with attraction strength parameters $\beta \epsilon_w = 0.2$, 0.5, 0.6, 0.72 and 0.74. The fluid-fluid particle interaction range is truncated to $L=5$.}
\label{fig:prof2d}
\end{figure}

Fig.~\ref{fig:prof2d} displays such 2D density profiles for external wall potentials of varying attraction strengths. These are calculated by solving Eq.\,\eqref{eq:difLatGas}. However, the value of $\mu$ is not at the outset imposed, instead the total adsorption \eqref{eq:adsorption} is specified. This is done by renormalising the density profile at each iteration of the algorithm via Eq.\,\eqref{eq:renorm}. For more details about this algorithm, see Ref.\ \onlinecite{hughes13}.

Fig.~\ref{fig:prof2d} shows that for weakly attracting walls (small $\epsilon_w$), it is energetically favourable for the liquid not to be in contact with the wall and the contact angle is large. As $\epsilon_w$ is increased, the contact angle decreases as the fluid seeks to have greater contact with the wall. As discussed below (see also Fig.~\ref{fig:bindingPot}), there is a wetting transition at $\beta \epsilon_w\approx 0.74$ and for values of $\beta \epsilon_w$ greater than this, the liquid spreads over the surface ($\theta=0$). {Note that the lattice-gas DFT, with various different choices for $\epsilon_{\boldi,\boldj}$, has been used extensively to study wetting and also to calculate profiles for liquid drops on surfaces -- see e.g.\ Refs.\,\onlinecite{detcherverry03, porcheron06, monson08} for some recent example. As far as we are aware, in all previous studies the adsorption on the surface is determined by the choice of $\mu$, either explicitly, or by fixing the total number of particles in the system, $N$. Since the fictitious potential that we use in our method to stabilise the drops is almost always rather small, the density profiles we obtain are actually rather similar to those found previously when the interaction and wall potentials are the same.}

\subsection{The Binding Potential via the Sharp-Kink Approximation}\label{sec:SK}
\begin{figure}
\begin{center}
\includegraphics[width=0.48\textwidth]{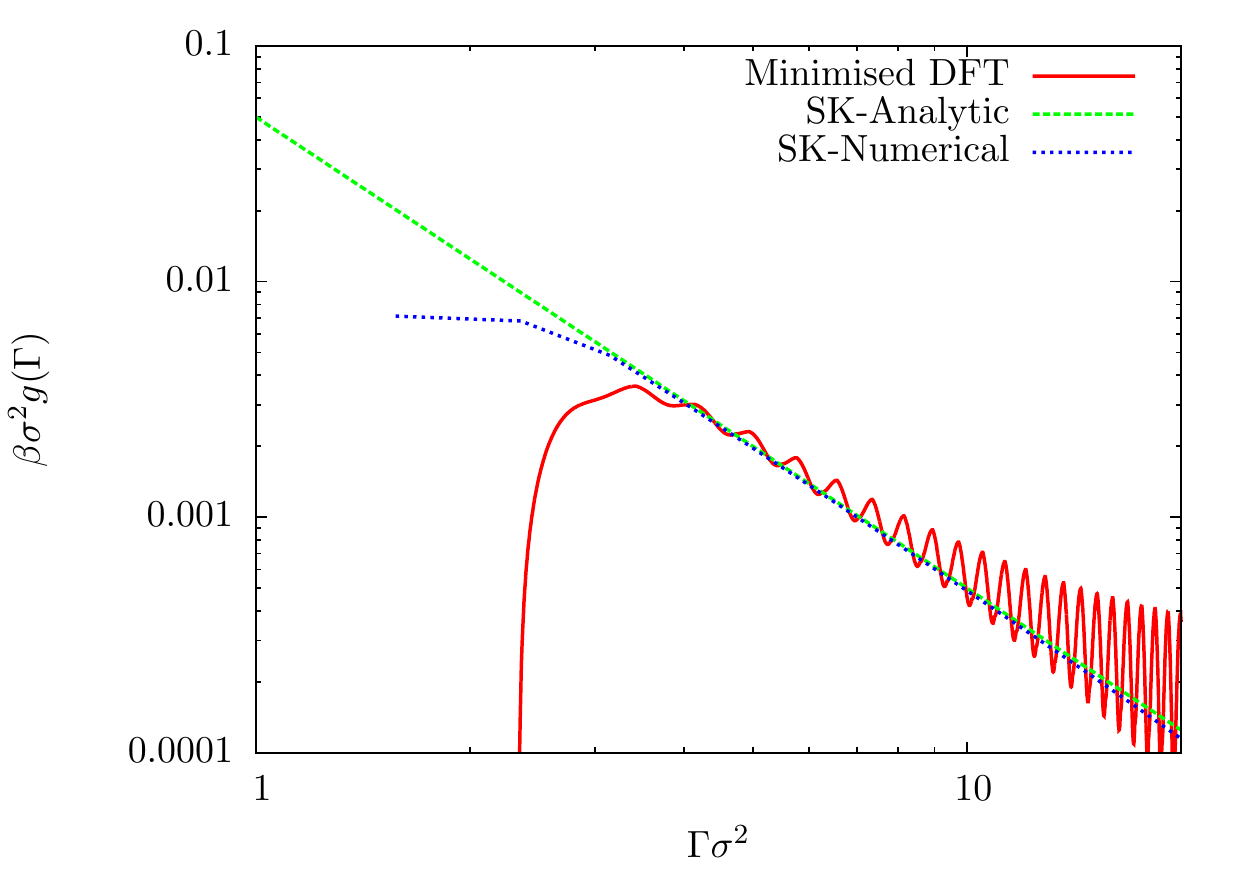}
\caption{The binding potential for $\beta \epsilon=0.9$ and $\beta \epsilon_w=0.55$ calculated for the lattice gas model and compared with results from the sharp kink approximation, displayed on a log-log plot. The green dashed line corresponds to the leading order ($\sim h^{-2}$) term in Eq.\,\eqref{eq:sk_asym} and the dotted blue line to a numerical evaluation of the sum in Eq.\,\eqref{eq:fullAsym}. The red solid line is the result from a constrained minimisation of the full functional.}
\label{fig:sklg_comp}
\end{center}
\end{figure}

Some of the interfacial and wetting behaviour of the lattice-gas model can be elucidated in a particularly simple manner by making the so-called sharp kink (SK) approximation. \cite{schick90,dietrich88} Specifically, the lattice gas density profile is taken to be
\begin{equation}
\rho_i = 
\begin{cases}
0 &\mbox{for } i \leq 0, \\
\rho_l &\mbox{for } 0 < i \leq h, \\
\rho_g &\mbox{for } i > h,
\end{cases}
\end{equation}
where $h$ is the position of the liquid-gas interface and the surface of the solid wall is at the lattice site $i=0$. No minimisation is necessary under this approximation and the binding potential for any given film thickness $h$ can be calculated directly. For a system with truncated fluid-fluid interactions, the binding potential can be calculated as
\begin{equation}\label{eq:fullAsym}
g(h) = (\rho_l- \rho_g){\sigma^3}\sum_{i=h+1}^\infty V_i,
\end{equation}
when $h>L$, the range of the fluid-fluid interactions. An asymptotic expansion of this sum gives (see also Ref.\,\onlinecite{macdowell14}):
\begin{equation}\label{eq:sk_asym}
g \sim \epsilon_w \pi (\rho_l- \rho_g){\sigma^3}\left(\frac{{\sigma^2}}{12h^2} + \frac{{\sigma^3}}{12h^3} + \cdots \right),
\end{equation}
Truncating this series gives an approximation that is valid for large $h$. The coefficients given in Eq.\,\eqref{eq:longAsym} can therefore be calculated explicitly for the lattice-gas model under this approximation. Figure \ref{fig:sklg_comp} shows the binding potential for $\beta\epsilon=0.9$ and $\beta\epsilon_w=0.55$ on a log-log plot, calculated using the full lattice-gas model and comparing with results from the SK approximation. The analytically calculated asymptotic limit in Eq.\,\eqref{eq:sk_asym} is plotted to $O(h^{-2})$, which agrees very well with a numerical evaluation of the full SK approximation. The numerical and analytic SK results begin to slightly drift apart for larger adsorptions. This is because in the numerical calculation the external potential is truncated to a range of $100\sigma$ whereas the analytic calculation assumes an infinite interaction range. This truncation results in the numerical result diverging from the $\sim h^{-2}$ decay in Eq.\,\eqref{eq:sk_asym} for large $h$.

The density profile obtained by minimising the DFT, in contrast to the SK approximation, incorporates a better approximation for the true shape of the liquid-gas interface and also the effect of this interface being close to the wall. Thus, for small values of $\Gamma$ (i.e.\ small $h$), this approximation is far more reliable. Also, it includes the correct $\Gamma^{-2}$ asymptotic decay for large $\Gamma$. Note that in Fig.\,\ref{fig:sklg_comp} the oscillations in the tail of the binding potential are present as a result of the system being discretised on a lattice. The free energy is lower when the liquid-gas interface is between two lattice sites, rather than on a lattice site. Thus, as the specified adsorption is increased, the liquid-gas interface moves continuously away from the wall which leads to oscillations in $g(\Gamma)$. These oscillations lie on top of the correct $\sim \Gamma^{-2}$ asymptotic decay; i.e.\ for large $h$ the binding potential is of the form $g \approx \epsilon_w \pi {\sigma^5}(\rho_l- \rho_g)/(12h^2)-B\cos(2\pi h/\sigma)$, where the amplitude of the oscillations $B$ is a small number that depends on the state point and manner in which the interactions are treated. As the range of the fluid-fluid interactions is increased, the amplitude of these oscillations $B$ decreases (c.f.\ Fig.\,\ref{fig:aPlot}). Thus, apart from these oscillations, the binding potential calculated from the full minimisation matches up well for large $\Gamma$ with the results from the SK approximation. {We attempted to extract the coefficients for the higher terms in the expansion in Eq.~\eqref{eq:longAsym} from our numerical results, since analytic expressions for these exist.\cite{dietrich91} However, the oscillations induced by the lattice in $g(h)$ make this problematic.} Note too that the free energy contribution from the liquid-gas free interface $\gamma_{lg}$ is a little different in the SK approximation compared to that obtained from the full minimisation. The respective values are subtracted when calculating $g(\Gamma)$ in the two methods.

For small values of the adsorption, large differences are seen between the SK results for $g(\Gamma)$ and those from the full minimisation. This demonstrates that the SK approximation should only be used for thicker films and a microscopic theory for the fluid structure (DFT) is required to accurately calculate a binding potential that is valid for thin ($h\lesssim 3\sigma)$ liquid films. 

\section{The Influence of the Range of Particle Interactions}\label{sec:ranges}
In Sec.\ \ref{sec:binding} the manner in which the particle interactions influence the (asymptotic) form of the binding potentials was discussed; i.e.\ for long range dispersion interactions the algebraic decay form in Eq.\,\eqref{eq:longAsym} is appropriate, whereas for short-range forces, the exponential form in Eq.\,\eqref{eq:shortAsym} is the correct asymptotic form. In this section, the cross-over from one regime to the other is explored as the interaction potentials in the lattice-gas model are truncated at different ranges.

When comparing systems with different interaction ranges, it is important that the truncation does not change the bulk fluid phase diagram. Otherwise, even if comparing systems at the same temperature $T$, the value of $(T-T_c)$, where $T_c$ is the bulk critical temperature, would not be the same for systems with a different truncation range $L$. For the simple mean-field approximation to the free energy in Eq.\,\eqref{eq:latGas}, the bulk uniform fluid free energy is determined by the integrated interaction strength of the pair potential $\sum \epsilon_{\boldi,\boldj}$, i.e.\ the total potential that arises from the interaction of a single particle with all others within the interaction range. As the interaction range is adjusted, one must vary the value of $\epsilon$, the parameter governing the overall strength of the pair interactions, to ensure that the integrated interaction strength remains constant so that the bulk fluid phase diagram remains unchanged. All values of $\beta \epsilon$ quoted here are the strength of the interaction when the interaction range is truncated to only the nearest neighbour lattice sites $(L=1)$.

\begin{figure}
\begin{center}
\includegraphics[width=0.48\textwidth]{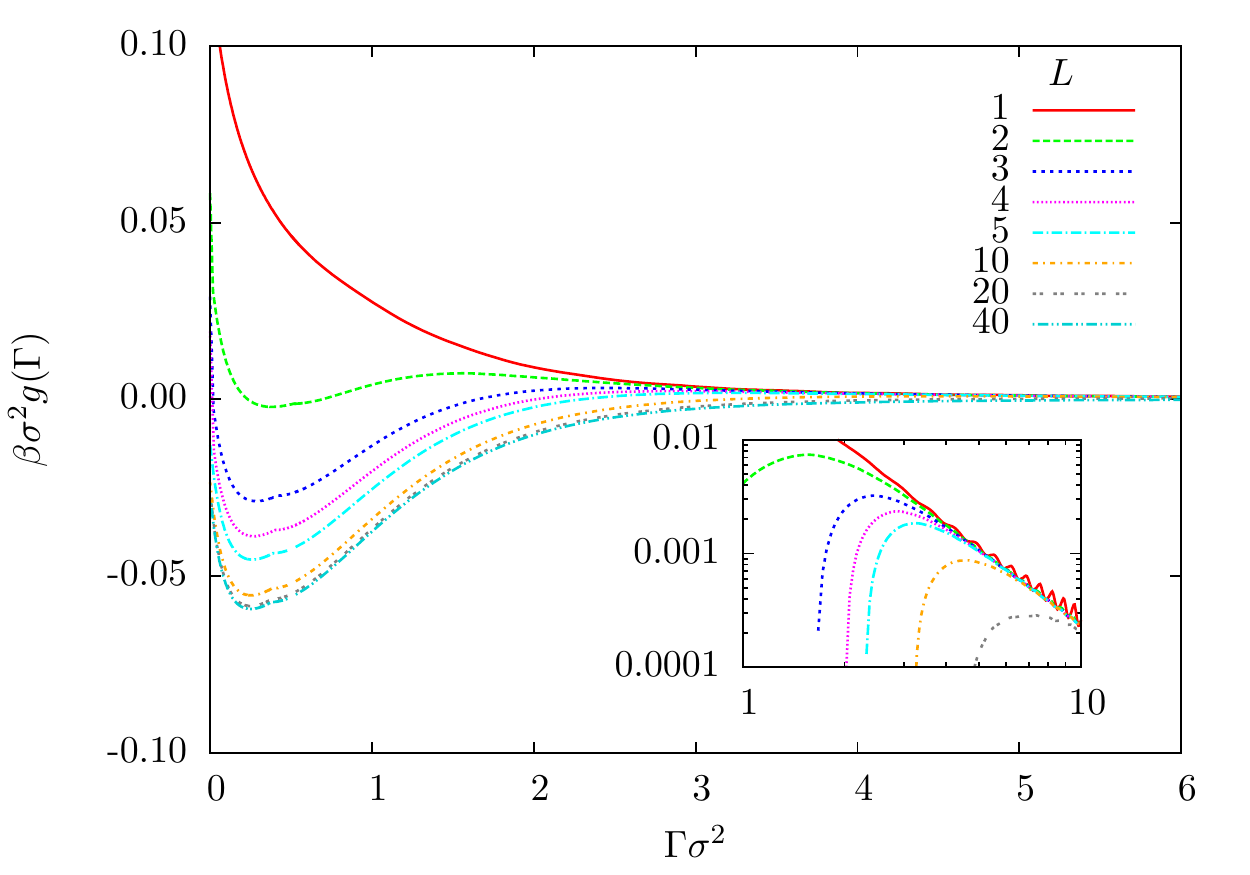}
\caption{The binding potential for the fluid with inverse temperature $\beta \epsilon=0.8$ at a wall with attraction strength $\beta \epsilon_w=0.5$. The various different curves correspond to truncating the fluid-fluid pair interactions at different values of the truncation length $L$. The wall-fluid interactions remain fixed at a truncation of $100\sigma$. Note that we vary $L$ in a manner that does not change the bulk fluid phase diagram. We see that the fluid is predicted to be more wetting as $L$ becomes shorter; in fact, the interfacial phase behaviour changes from non-wetting to wetting, purely as a result of truncating the interaction range. The inset shows the same curves plotted on a log-log scale which shows that the asymptotic decay form for large $\Gamma$ is the same in all cases because of the long-ranged wall-fluid interactions.}
\label{fig:aPlot}
\end{center}
\end{figure}

\begin{figure}
\begin{center}
\includegraphics[width=0.48\textwidth]{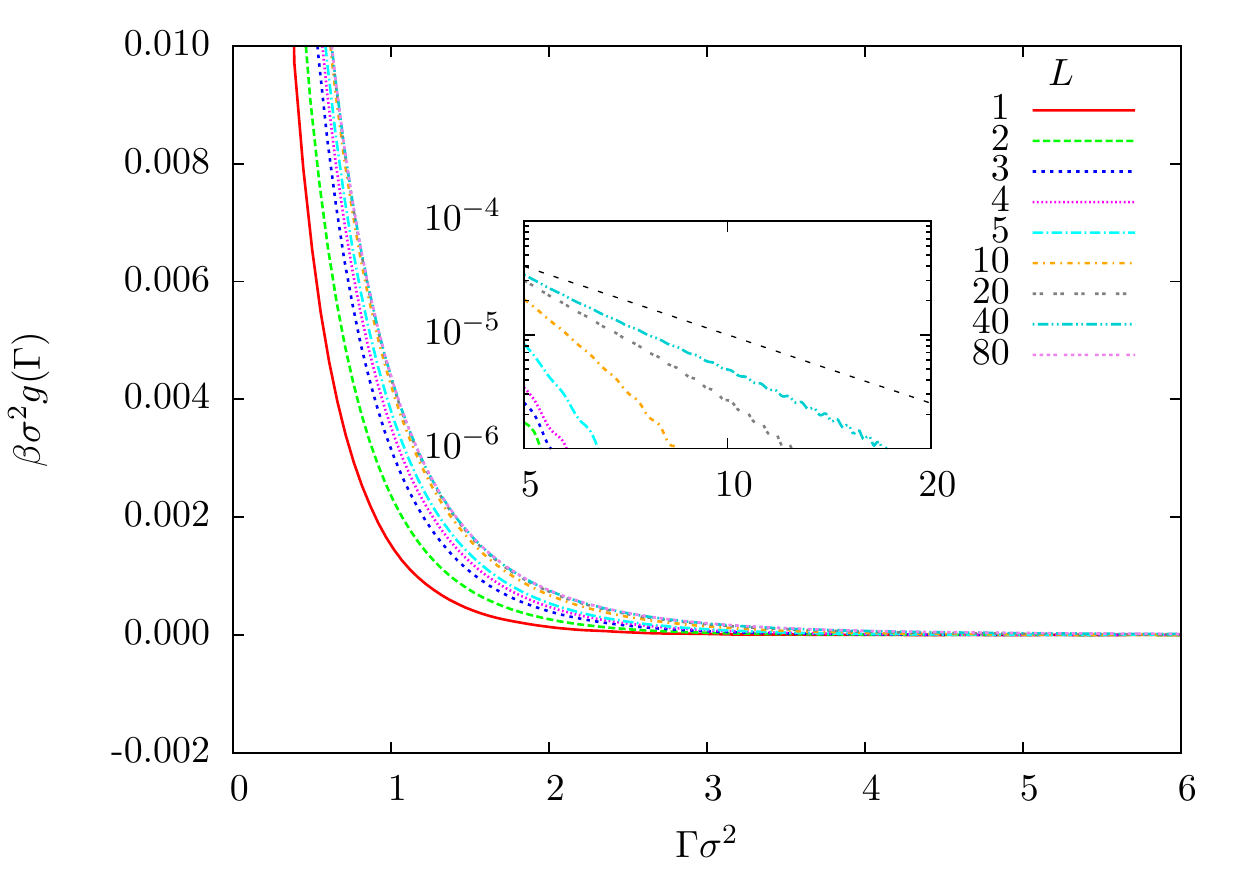}
\caption{The binding potential for the fluid with inverse temperature $\beta \epsilon=0.9$ at a wall with attraction strength $\beta \epsilon_w=0.7$. For these values, the fluid wets the wall. In contrasts to the case in Fig.~\ref{fig:aPlot}, here we truncate both the wall-fluid and fluid-fluid interactions at the same range, $L$. The different curves correspond to varying $L$ from between 1 to 80 particle diameters. As $L$ is decreased, the range of $g(\Gamma)$ decreases. In the inset, the black dashed line is $0.001\Gamma^{-2}$. As $L$ is increased, the binding potential approaches this large $\Gamma$ asymptotic decay form.}
\label{fig:bPlot}
\end{center}
\end{figure}

Truncating the range over which particles in the system interact changes the overall shape of the resulting binding potential. In particular, if all interaction potentials (both fluid-fluid and wall-fluid) are truncated, then the tail of the binding potential decays exponentially to zero. If there are any long ranged (not truncated) interactions then, in three dimensions, the binding potential tail decays algebraically $\sim h^{-2}$. By varying the range of the particle interactions in the lattice-gas model, both of these regimes are seen and also there is a crossover from one to the other as the truncation range $L$ is varied. These results are displayed in Figs.~\ref{fig:aPlot} and \ref{fig:bPlot}.

Fig.~\ref{fig:aPlot} shows the binding potential for $\beta\epsilon=0.8$, $\beta\epsilon_w=0.5$ and at different truncation ranges, $L$, of the fluid-fluid interactions. Note, however, that the interactions between wall and fluid particles extends over the entire domain in all cases (the domain size is $M_k=100\sigma$). As the truncation range $L$ is increased, it becomes energetically favourable for the fluid not to wet the substrate and so a minimum in $g(\Gamma)$ at a finite value of the adsorption $\Gamma$ appears; i.e.\ the interfacial phase behaviour changes purely as a result of how the fluid-fluid particle interactions are modelled. Note that since the value of $g(\Gamma)$ at the minimum $g(\Gamma_0)\equiv g(h_0)$ decreases as $L$ increases, from Eq.\,\eqref{eq:bindCont}, this indicates that increasing the interaction range $L$ makes the fluid less wetting and increases the contact angle. This shows that care should always be taken when modelling the interaction between two fluid particles: truncating the pair potentials at too small a distance may result in significant errors in predictions for the interfacial phase behaviour. Fig.~\ref{fig:aPlot} also shows the $\sim \Gamma^{-2}$ decay of $g(\Gamma)$ as $\Gamma\to\infty$ that is present in all cases, due to the presence of the long ranged wall-fluid interactions.

The effect of truncating the range of all interactions (including the wall-fluid potential) on the binding potential is shown in Fig.\,\ref{fig:bPlot} for the case when $\beta \epsilon=0.8$ and $\beta \epsilon_w=0.7$. The binding potential for $L=80$ is the longest ranged, having the slowest decay to zero as $\Gamma$ increases. As $L$ is decreased, the range of $g(\Gamma)$ decreases. The inset of the figure shows the same data on a log-log scale which enables one to clearly see the form of the asymptotic decay of $g(\Gamma)$ as $\Gamma \to \infty$. As expected from the discussion above, as the range of the interactions $L$ is increased, an increasingly large portion of algebraic $\sim\Gamma^{-2}$ decay is present. However, it should be pointed out that formally speaking, it is only when $L\to\infty$ that the ultimate asymptotic decay of $g(\Gamma)$ as $\Gamma \to\infty$ changes from exponential to algebraic.

\section{A Fitting Function for the Binding Potential}\label{sec:fitting}
\begin{figure}
\begin{center}
\includegraphics[width=0.48\textwidth]{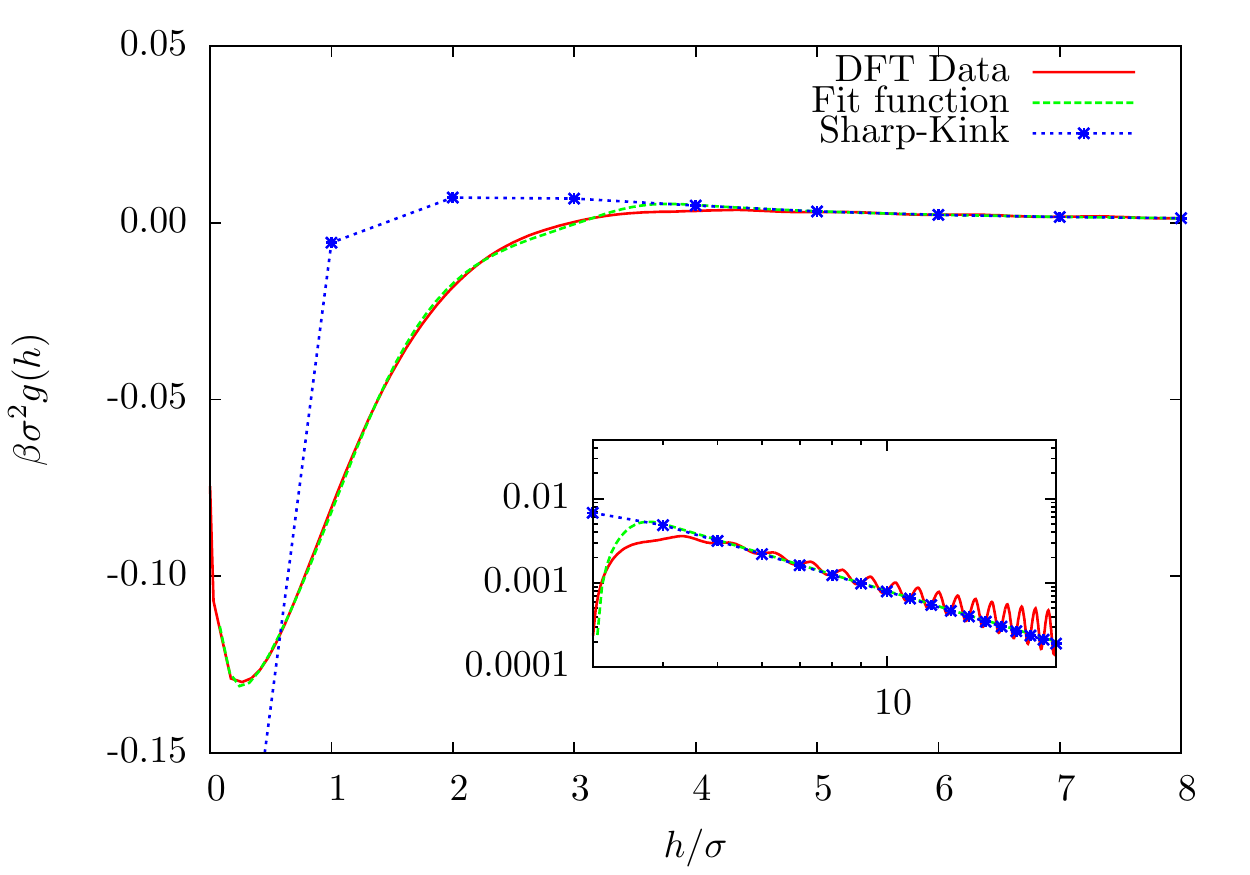}
\caption{The binding potential for a fluid with $\beta \epsilon=0.9$ and $\beta \epsilon_w=0.55$ with truncated interaction ranges of $5\sigma$ and $100\sigma$, for the fluid-fluid and wall-fluid interactions, respectively. The data calculated from the DFT model is shown with the fitting function Eq.\,\eqref{eq:expFit} (parameter values are given in {Appendix A}) and the sharp-kink binding potential for the same parameters. The inset shows the same data on a log-log scale.}
\label{fig:sk_comp}
\end{center}
\end{figure}

In order to take the binding potentials calculated via DFT in the previous section and use them with the mesoscale IH model, a suitable fit function is required. The fit function should have the same form as that observed in the DFT results so that it can accurately fit the data. Several aspects of the binding potential curves are particularly important for the fit function to be correct: At small values of the adsorption $\Gamma$, the binding potential exhibits a minimum and a maximum when the fluid is non-wetting or near to the wetting transition. These need to be fitted well; in particular the value at the minimum $g(h_0)$ needs to be the correct value [c.f. Eq.\,\eqref{eq:bindCont}]. Also, the true binding potential is finite for $\Gamma=0$ (unlike the approximation in Eq.\,\eqref{eq:longAsym} and other such power-series) and so the fit function should not diverge at $\Gamma=0$. Finally, when dispersion interactions are present, the fit function should exhibit the correct $\Gamma^{-2}$ decay as $\Gamma\to\infty$. Therefore, the following form for the fit function is suggested:
\begin{equation}\label{eq:expFit}
g(\Gamma) = A \frac{\exp[- P(\Gamma)] -1}{\Gamma^2},
\end{equation}
with
\begin{equation}
P(\Gamma) = a_0 \Gamma^2 e^{-a_1 \Gamma} +a_2 \Gamma^2 + a_3 \Gamma^3 + a_4\Gamma^4 + a_5\Gamma^5.
\end{equation}
The rationale for this choice is as follows: For small $x$, $\exp(x) \approx 1+x$, and so at low adsorptions this form gives $g(\Gamma) \approx A(a_0 e^{-a_1 \Gamma} +a_2 + a_3 \Gamma + \dots)$. For high adsorptions this form gives $g(\Gamma) \sim -A \Gamma^{-2}$, which is the correct form for the asymptotic $\Gamma\to\infty$ decay. The coefficient $a_5$ of the highest order term must be positive. The second exponential within the outer exponential function [i.e.\ the exponential term with coefficient $a_0$ in $P(\Gamma)$] helps to correctly fit the minimum at low adsorptions, which is often asymmetric, being much steeper on the low adsorption side of the minimum, compared to the other side, as can seen in Fig.\,\ref{fig:sk_comp}. The constant $a_1$ in Eq.\,\eqref{eq:expFit} is usually quite large and positive so that the inner exponential term has almost no effect on the form of $g(\Gamma)$ on the large adsorption side of the minimum.

An example of the fitting function is displayed in Fig.\,\ref{fig:sk_comp}. The function is plotted with the original data and the SK results as a comparison. The SK results do not describe the behaviour at small $\Gamma$ whereas the fit function gives a very good approximation to the data over the whole range. The leading order coefficient of the decay of the binding potential, $A$ in Eq.\,\eqref{eq:expFit}, can either be calculated directly using the SK approximation or can be found by fitting the DFT data, both methods give similar results. The value for $A$ from the SK approximation is used here. In {Appendix A} we give the fit parameter values in Eq.~\eqref{eq:expFit} for a range of values of $\beta\epsilon$, $\beta\epsilon_w$ and $L$.

\section{Droplet Profiles Obtained Using the Binding Potential}\label{sec:results}
Using the binding potential calculated from the DFT, a drop thickness profile can be calculated from the IH model, by minimising the free energy in Eqs.\,\eqref{eq:fullcurveenergy} or \eqref{eq:thinEnergy}. This drop profile, despite being the result of a mesoscale calculation, contains information about the nature of the microscopic interactions between particles in the system via the binding potential. Of course, a drop profile can also be calculated directly using DFT; the result of such a calculation is shown in Fig.\,\ref{fig:prof2d}. However, it is computationally easier to treat larger systems using the IH model. Also, non-equilibrium situations are much more easily modelled via Eq.\,\eqref{eq:thinFilm} than with a dynamical DFT model that includes all the hydrodynamics.\cite{archer06, archer09} However, since the two approaches are both based upon the same microscopic interactions, the resulting drop profiles from each method should be the same at the mesoscopic scale. This section shows how the two approaches compare. {We find that overall, the drop shape profiles from the DFT and from the IH model are in good agreement, a result which {\em a-priori} is not obvious, if bearing in mind the degree of coarse-graining in going from a density profile to a film height profile. The good agreement between the two is thus more than a mere consistency test.}

\begin{figure}
\vspace{-3em}
\includegraphics[width=0.48\textwidth]{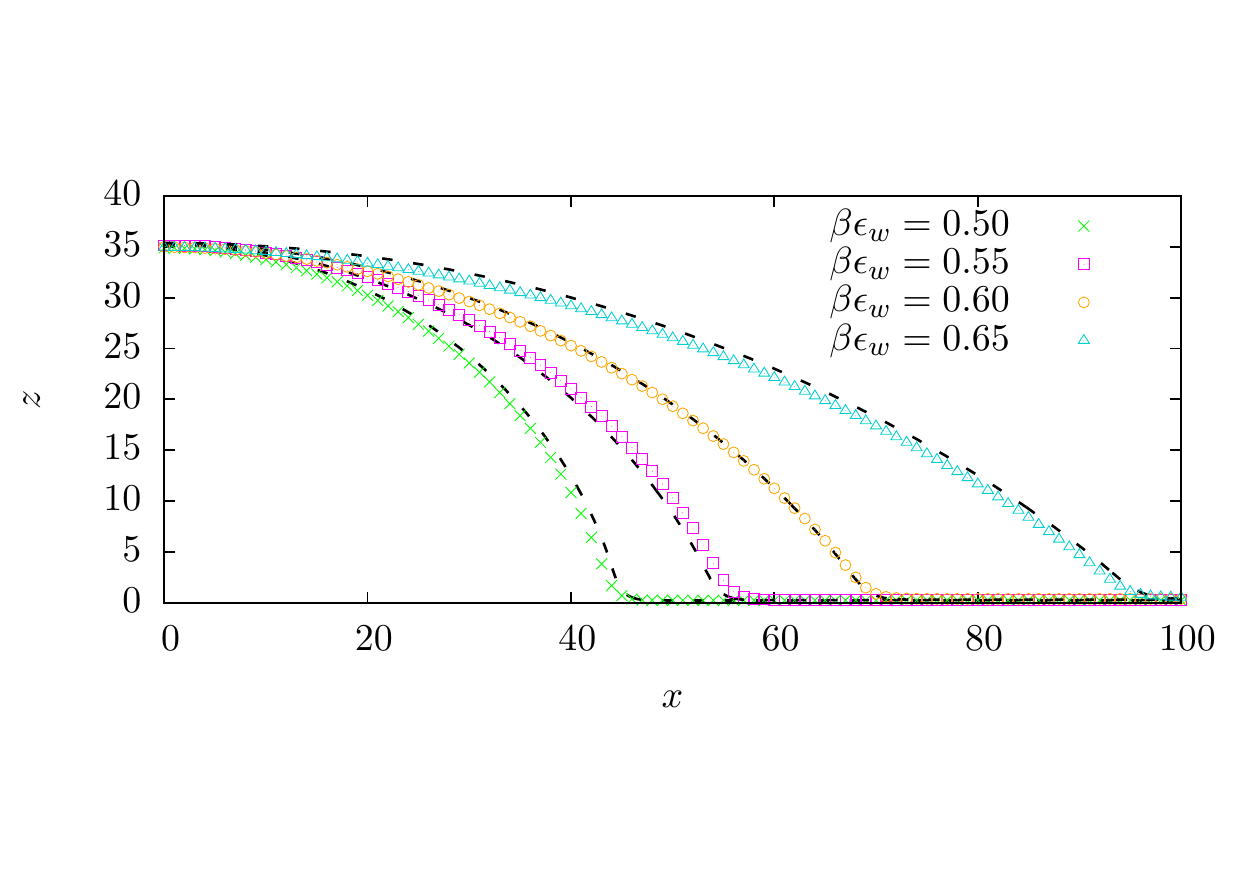}

\vspace{-3em}
\caption{Comparison of drop profiles calculated from DFT (points) and from the IH model, Eq.\,\eqref{eq:fullcurveenergy} (dashed lines). We compare drops of equal maximum height for a range of values of the wall attraction strength $\epsilon_w$. The two methods coincide well across the range of contact angles. These are for the case when the particle interactions are truncated to a range of $L=5$ with a strength of $\beta \epsilon=0.9$. The parameters used in the fitting functions for $g(h)$, used to obtain the IH results, can be found {in Appendix A}.}
\label{fig:drops}
\end{figure}

\begin{figure}
\includegraphics[width=0.48\textwidth]{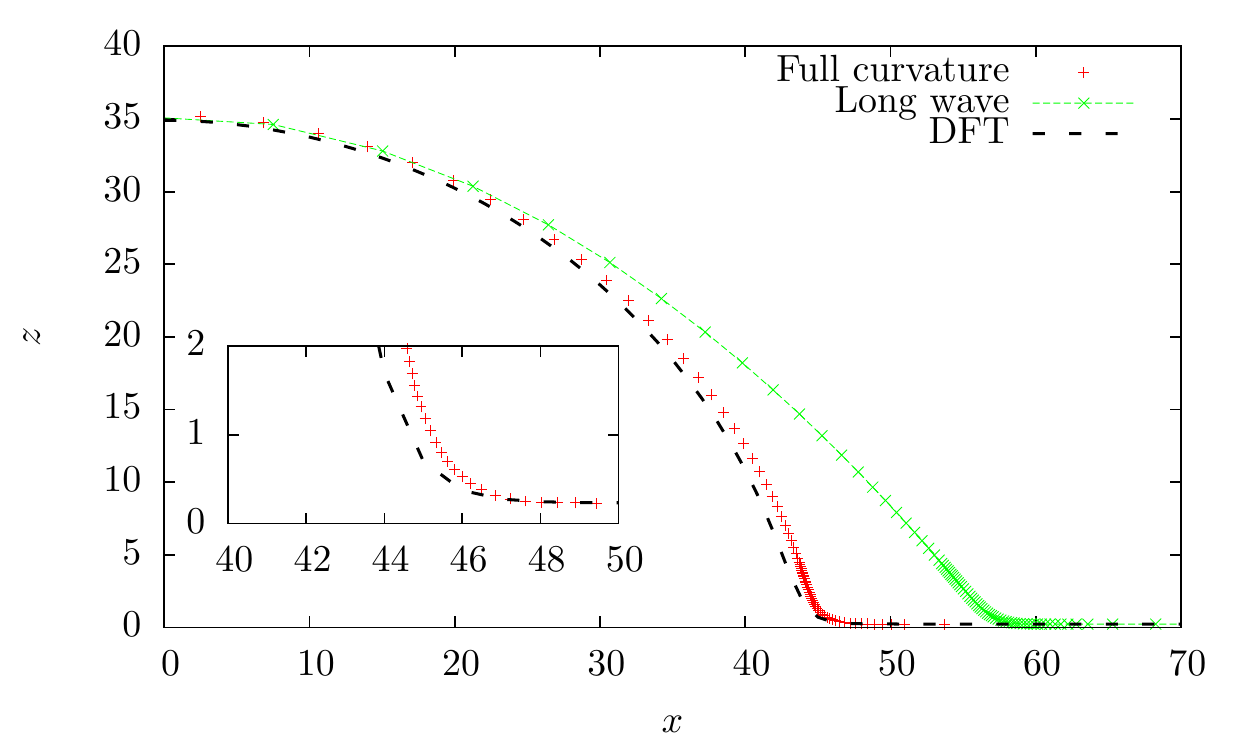}
\includegraphics[width=0.48\textwidth]{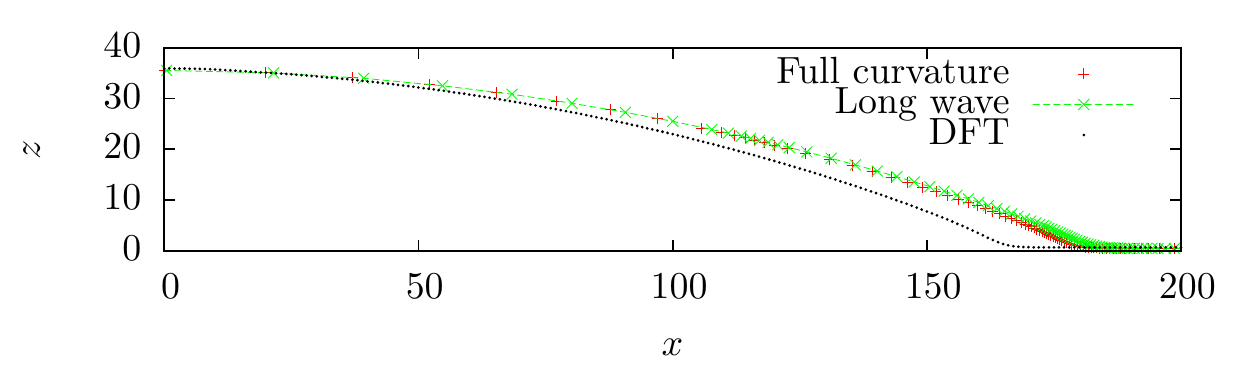}
\caption{Drop profiles calculated from the DFT density profiles together with those from the IH model using both the full curvature free energy \eqref{eq:fullcurveenergy}, and the long wavelength approximation, Eq.\,\eqref{eq:thinEnergy}. These results are for $\beta\epsilon=0.9$ and $\beta\epsilon_w=0.5$ (top figure) and $\beta\epsilon_w=0.7$ (bottom figure). In the inset we show a magnification of the contact line region. Note the `precursor' film height $\ll \sigma$ indicating that this is sub-monolayer.}
\label{fig:full/longwave}
\end{figure}

Performing the minimisation described in Sec.\ \ref{sec:lattice} for the lattice-gas DFT{, constrained via Eq.\,\eqref{eq:renorm},} on a 2D domain yields density profiles such as those displayed in Fig.\,\ref{fig:prof2d}. These 2D profiles correspond to the density profile of a cross section through a liquid ridge on a surface with centre of mass along the line $x=0$. The location of the liquid-gas interface can be calculated using Eq.\,\eqref{eq:filmHeight}. The alternative procedure, which yields a very similar result, is to just plot the contour where the density $\rho_\boldi=(\rho_g+\rho_l)/2=0.5$. Note that this treats the (strictly) discrete density profile as a continuous function. In Fig.\,\ref{fig:drops} are displayed drop profiles obtained from DFT and using Eq.\, \eqref{eq:filmHeight}, compared with results obtained from minimising Eq.\,\eqref{eq:fullcurveenergy} together with the binding potential obtained from DFT for various values of $\beta \epsilon_w$. These results show that drop profiles obtained via the two methods coincide well with each other, {as was also observed in Ref.~\onlinecite{nold14}}. In Fig.\,\ref{fig:full/longwave} are comparisons of a droplet profile found via the DFT route with droplets calculated from a minimisation of both the full curvature free energy Eq.\,\eqref{eq:fullcurveenergy} and the long wavelength approximation free energy, Eq.\,\eqref{eq:thinEnergy}. This shows that the full curvature free energy \eqref{eq:fullcurveenergy} gives closer agreement with the DFT results than the approximation in Eq.\,\eqref{eq:thinEnergy}, as one would expect. Note too that when the contact angle is very small, then the drop shape is very sensitive to variations in the parameters. 

Determining the contact angle from drop profiles obtained using the IH model is not straight-forward because the profiles have a precursor film. The DFT calculations show (see e.g.\ Figs.\,\ref{fig:profs}, \ref{fig:prof1d} and \ref{fig:prof2d}) positive adsorption on the surface outside of a droplet. However, the fluid density in the first ($i=1$) layer of lattice sites is substantially less than one -- see also the inset in Fig.\,\ref{fig:full/longwave}. This indicates that the density on the surface outside of the drops is not even a complete monolayer and so calling it a ``precursor film'' is arguably misleading.

Returning to the issue of calculating the contact angle in a IH pre-cursor film model: here, this is calculated from the curvature at the top of the drop, at $x=0$. Liquid droplets of a suitable size have the shape of a spherical cap. This is the case when the volume of the drop is small enough that gravity does not play a significant role but also large enough that the shape at the maximum is not distorted by the binding potential. It is droplets of this size that are studied here. In these cases where the drops have this spherical cap shape, the contact angle for the drop is calculated by fitting a circle to the highest point of the drop profile. The contact angle that this circle makes with the substrate is taken to be the contact angle of the drop. This procedure is straight-forward: The curvature at a stationary point, i.e.\ the maximum point of the liquid drop, is simply the second derivative of the height profile at that point and so the radius of curvature is given as
\begin{equation}
r_c= \frac{1}{h''(x_\text{max})}
\end{equation}
where $x_\text{max}$ is the point where the height of the droplet $h$ is at it's maximum value. Defining $a$ to be the distance from the point that the circle meets the substrate to the centre of the base of the drop, then the contact angle is given by
\begin{equation}
\theta_c = \frac{\pi}{2}- \cos\left(\frac{a}{r_c}\right).
\end{equation}

\begin{figure}
\begin{center}
\includegraphics[width=0.48\textwidth]{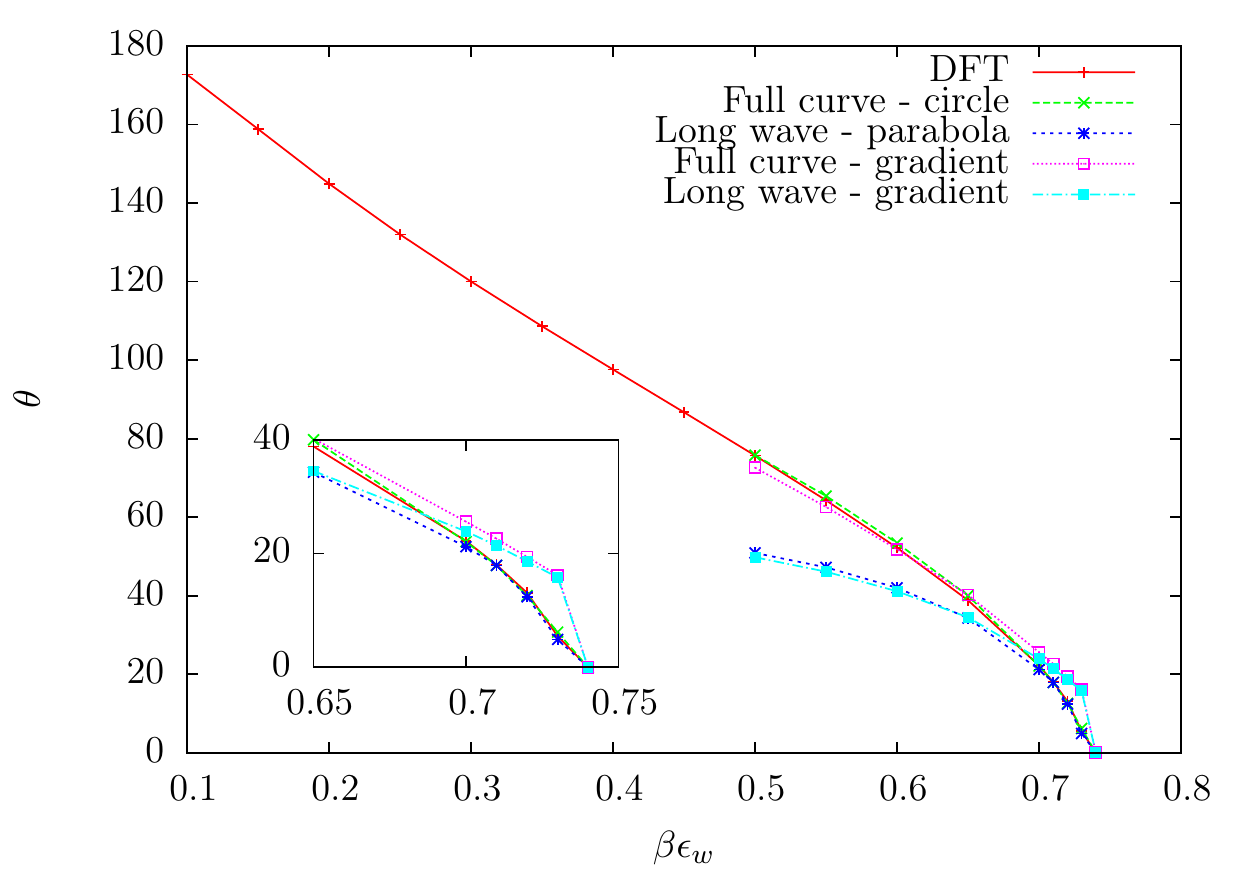}
\caption{The contact angle for a range of different values of $\beta \epsilon_w$, for $\beta\epsilon=0.9$, calculated from both the DFT and IH models. Contact angles are calculated for both the full curvature and long-wave approximation versions of the IH model, using both the circle fitting and maximum gradient methods. The circle fitting method does not well represent the actual contact angle of the droplet from the long-wave IH model.}
\label{fig:cntAngle}
\end{center}
\end{figure}

Comparisons made in this way can only be used to relate the two theories for systems where a liquid drop exists, i.e.\ for a partially wetting system. The focus here is also on drops with an acute contact angle as these profiles can be calculated by minimising Eq.\,\eqref{eq:thinEnergy}, but droplets with obtuse contact angles cannot. For non-equilibrium problems one makes the assumption of small contact angles (long-wavelength) and utilises the thin film equation, Eq.~\eqref{eq:thinFilm}, with the free energy functional given by Eq.\,\eqref{eq:thinEnergy}. We should also mention that making the long-wave approximation, i.e.\ going from Eq.~\eqref{eq:fullcurveenergy} to Eq.~\eqref{eq:thinEnergy}, also results in the drop shape away from the surface ceasing to be the arc of a circle and instead being a parabola.\cite{safran94} Thus, for the long-wave theory \eqref{eq:thinEnergy}, we fit the drops with a parabola, in order to make a fair comparison. Fig.~\ref{fig:cntAngle} shows the comparison of contact angles calculated for different values of $\beta \epsilon_w$, from the DFT model and from the IH model, for both the full curvature case and the long-wavelength approximation.

An alternate way to obtain the contact angle is by calculating the largest gradient of the drop profile. The inverse tangent of this gives the contact angle. These `maximum gradient' contact angles are also shown in Fig.\,\ref{fig:cntAngle}. This figure shows that results based on the full curvature free energy agree well with those calculated from the DFT model. The IH model in the long-wave approximation only agrees with the DFT results for very small contact angles, as expected. For the full curvature model, the contact angle found by fitting a circle to the profile matches the DFT results better than those from the maximum gradient method. The droplet shape calculated under the long-wavelength approximation is no longer that of a spherical cap, it is instead parabolic, and so a similar procedure to the circle fitting method is employed where a parabola is fitted to the droplet profile. Results for the contact angle obtained via these fits are shown in Fig.\,\ref{fig:cntAngle}. Clearly, these results depend heavily on the method used to find the contact angle of the droplets and although the circle/parabola fitting method gives better results it is only applicable to specific drop shapes -- see Ref.~\onlinecite{tretyakov13} for a further discussion on extracting a contact angle from a drop profile. Note also that the contact angles in Fig.\,\ref{fig:cntAngle} calculated via DFT are the only results to extend beyond $90^\circ$; droplet profiles can not be found in this range using the IH model.

The calculated contact angles {are weakly dependent on the size of the drop}. All contact angles {extracted from the IH model are calculated} from droplets that have a height of $35\sigma$. Calculating contact angles from droplets of a specific height, rather than a specific volume, seems to give more consistent results. {The range at which particle interactions are truncated does not significantly affect how well the results from the two models (DFT an IH) agree with each other. Over the range of parameters studied, up to a truncation range of $L=40$, the discrepancy in the contact angle between the two models is typically a few percent.}

\section{Conclusions}\label{sec:conclusion}

In this paper we have developed a microscopic DFT based method for calculating the binding potential $g(\Gamma)$ at various different state points and interaction potentials of various different strengths and ranges between the surface and the liquid. These were subsequently used as input to a mesoscopic  {interface} free energy model, which was then used to calculate the height profile of liquid drops on surfaces. The liquid height profiles are very similar in shape to the profiles obtained directly from the DFT, indicating that the coarse-graining procedure used to obtain $g(\Gamma)$ is valid.

The binding potential is calculated as a function of the adsorption $\Gamma$, which can be related to the film height via Eq.\ \eqref{eq:adsorption_approx}. It is calculated using the method developed in Ref.\ \onlinecite{archer11} for studying nucleation of the liquid from the gas phase. This method constrains the fluid density profile to have a specified value of $\Gamma$, which is equivalent to imposing an additional fictitious external potential that stabilises the adsorbed film with the given adsorption. Note that this fictitious potential is not known a-priori and is calculated on-the-fly as part of the minimisation to obtain the fluid density profile.

The method presented for calculating the binding potential is general and it should be possible to use if for any DFT. Here it has been implemented using a simple DFT for a lattice-gas, that in Eq.\ \eqref{eq:latGas}. A better description of the free energy of a lattice-gas is possible, we refer the interested reader to e.g.\ Ref.\ \onlinecite{edison14b} for a more sophisticated and accurate approximation for the lattice-gas free energy. The lattice-gas approximation was used in order to more easily test the results for a range of different drop sizes. Note that to calculate the 2D density profiles for drops of the size in Fig.\ \ref{fig:prof2d} (or even larger) with a more sophisticated continuum DFT can be computationally time consuming -- e.g.\ using the commonly used approximation
${\cal F}[\rho({\bf r})]={\cal F}_{FMT}[\rho({\bf r})]+{\cal F}_{att}[\rho({\bf r})]$, where ${\cal F}_{FMT}[\rho({\bf r})]$ is the fundamental-measure theory approximation of the free energy of a fluid of hard-spheres and ${\cal F}_{att}[\rho({\bf r})]$ is a simple mean-field approximation for the contribution to the free energy due to the attractive interactions between the fluid particles.\cite{evans92,hansen13,Lutsko10,Tarazona08,evans10,lowen10} However, one of the draw-backs of the lattice-gas model is that the discretization onto a lattice leads to unrealistic small amplitude oscillations in $g(\Gamma)$, particularly for larger values of $\beta \epsilon$ (i.e.\ for lower temperatures) -- see e.g.\ Fig.\ \ref{fig:sklg_comp}. A more realistic continuum DFT model will not exhibit such oscillations that do not decay in amplitude with distance from the wall. Note, however, that oscillations may be present in the true binding potential. These, if present, stem from the packing of the particles, rather than from the discretization onto a lattice and are to be expected whenever the fluid exhibits layering transitions.

{Another drawback of the lattice-gas is that the mapping of the continuum fluid onto the lattice constitutes a significant approximation. This prevents a quantitative comparison between our results with existing simulation results for continuum models, such as those in Refs.\ \onlinecite{macdowell05, macdowell06, herring10, macdowell11, rane11, gregorio12, tretyakov13, benet14}. However, since we do find good qualitative agreement, we are currently applying the method using continuum (FMT) DFT. Results from this follow-on study will be published elsewhere, including comparison with simulation results.}

It should also be emphasised that it remains straight-forward to calculate droplet profiles using DFT rather than using the mesoscale model over the full range of contact angles, i.e.\ including when the contact angle $>90^\circ$. However, the IH model is particularly advantageous when considering the non-equilibrium situation, which can be described using Eq.\,\eqref{eq:thinFilm}, albeit this equation is derived under the assumption of small contact angles and only accounts for convective transport with no slip at the substrate. It is shown in Fig.\,\ref{fig:cntAngle} how the accuracy rapidly breaks down for larger contact angles. Non-equilibrium phenomena can also be studied using dynamical DFT, see e.g.\ Refs.\ \onlinecite{marconi99, marconi00, archer04a, archer04b} for more on this approach. 

One of the striking results of the present work is the observation of the degree to which the binding potential changes when the range at which particle interactions are truncated is changed -- see in particular Figs.\ \ref{fig:aPlot} and \ref{fig:bPlot}. Crucially, the value of $g(\Gamma)$ at the minimum that can be present at low values of the adsorption depends very sensitively on the truncation range. This minimum can change from being just a local minimum (a metastable equilibrium point) to become the global energetic minimum which represents a shift in the phase behaviour from wetting to non-wetting. In fact, for very short truncation ranges, the minimum can be completely removed. The conclusion from this part of our study is that to determine accurately the location of a wetting transition in theory or simulations one must ensure that the truncation is not so severe as to induce such errors. The tails of the potentials matter!

{This is important in the context of (coarse-grained) Molecular Dynamics simulations where dispersion forces are often cut off at distances of a few particle diameters. Ref.~\onlinecite{tretyakov13}, for instance, employs a cut-off length for the Lenard Jones interactions of two times the equilibrium bead distance and finds, in consequence, that the extracted binding potential is well fitted by a sum of exponentials as in Eq.~(\ref{eq:shortAsym}), similarly to our results when only short-range interactions are considered.}

Note also that for weakly attracting (solvophobic) walls the minimum in $g(\Gamma)$ is at very small, or potentially even negative, values of $\Gamma$. In this regime describing the sub-monolayer adsorption via a film-thickness $h$ is a somewhat misleading concept. In this regime, the minimum in the binding potential is very asymmetric, rising very sharply on the small $\Gamma$ side of the minimum, but rising far less steeply as $\Gamma$ is increased from the value at the minimum. To incorporate this behaviour in the fit-function for $g(\Gamma)$, we had to use the exponential in the function $P(\Gamma)$ in Eq.\ \eqref{eq:expFit}. This also means that the minimum in $g(\Gamma)$ is not well-approximated by a quadratic function when the minimum is at a small value of $\Gamma$ and so, of course, capillary-wave theory also does not apply in this regime, since capillary-wave theory assumes Gaussian fluctuations in a harmonic potential. Also, when the adsorption at the surface is sub-monolayer, then a non-equilibrium situation can no longer be described by Eq.\ \eqref{eq:thinFilm}; it can instead be described via a gradient dynamics model with a diffusive dynamics.

Finally, it should also be pointed out that although in the present work we have developed a method for calculating a better approximation for the binding potential, our approach is still a mean-field theory and therefore does not include all fluctuation effects, because our theory is based on Eq.\ \eqref{eq:fullcurveenergy}. Some effects of fluctuations at wetting transitions can be included by replacing $\gamma_{lg}$ in Eq.\ \eqref{eq:fullcurveenergy} with a function $\gamma(h)$. However, as Parry and co-workers have showed, the {effective interface} Hamiltonian is in fact non-local.\cite{parry06, parry07, bernardino09} Non-locality and fluctuation effects are particularly important at wetting transitions.

\section*{Acknowledgements}

A.P.H. acknowledges support through a Loughborough University Graduate School Studentship. All authors thank the Center of Nonlinear Science (CeNoS) of the University of M\"unster for recent support of the authors collaboration.

\begin{appendix}
\section{Fit function parameters}\label{app:param}
The following table contains the parameters to the fitted binding potentials that have been used throughout this paper {and also for a selection of other state points}. For ease of reference, the fitting function is repeated here:
\begin{equation}
g(\Gamma) = A \frac{\exp[- P(\Gamma)] -1}{\Gamma^2},
\end{equation}
with
\begin{equation}
P(\Gamma) = a_0 \Gamma^2 e^{-a_1 \Gamma} +a_2 \Gamma^2 + a_3 \Gamma^3 + a_4\Gamma^4 + a_5\Gamma^5.	\nonumber
\end{equation}

\begin{table*}
\begin{tabular}{c>{$}c<{$}>{$}c<{$}>{$}c<{$}>{$}c<{$}>{$}c<{$}>{$}c<{$}>{$}c<{$}>{$}c<{$}>{$}c<{$}>{$}c<{$}>{$}c<{$} }
Figure & \beta \epsilon & \beta \epsilon_w & L & A & a_0 & a_1 & a_2 & a_3 & a_4 & a_5 & a_6\\[1em] \hline
\ref{fig:drops} \& \ref{fig:full/longwave}&  0.9 & 0.5 & 5 & -0.073 & 1.142 & 8 & -3.078 & 3.283 & -1.272 & 0.173 & 0.000 \\
\ref{fig:drops} \& \ref{fig:sk_comp} & 0.9 & 0.55 & 5 & -0.081 & 1.080 & 8 & -2.153 & 2.074 & -0.705 & 0.084 & 0.000 \\
\ref{fig:drops} &  0.9 & 0.6 & 5 & -0.088 & 0.964 & 8 & -1.325 & 0.930 & -0.036 & -0.097 & 0.019 \\
\ref{fig:drops} &  0.9 & 0.65 & 5 & -0.096 & 0.770 & 8 & -0.534 & -0.245 & 0.762 & -0.358 & 0.051 \\
\ref{fig:full/longwave} & 0.9 & 0.7 & 5 & -0.103 & 0.609 & 8 & 0.138 & -1.037 & 1.158 & -0.451 & 0.060 \\
- & 0.8 & 0.5 & 2 & -0.062 & 0.431 & 8 & 0.431 & -1.448 & 1.562 & -0.649 & 0.097 \\
- & 0.8 & 0.5 & 3 & -0.062 & 0.562 & 8 & -0.203 & -0.900 & 1.175 & -0.468 & 0.062 \\
- & 0.8 & 0.5 & 4 & -0.062 & 0.669 & 8 & -0.473 & -0.548 & 0.903 & -0.363 & 0.047 \\
- & 0.8 & 0.5 & 5 & -0.062 & 0.776 & 8 & -0.679 & -0.222 & 0.634 & -0.262 & 0.033 \\
- & 0.8 & 0.5 & 10 & -0.062 & 0.998 & 8 & -1.035 & 0.415 & 0.107 & -0.074 & 0.009 \\
- & 0.8 & 0.5 & 20 & -0.062 & 1.125 & 8 & -1.194 & 0.752 & -0.162 & 0.011 & 0.0001 \\
- & 0.8 & 0.5 & 40 & -0.062 & 1.104 & 8 & -1.179 & 0.678 & -0.100 & -0.007 & 0.002 \\
- & 0.9 & 0.6 & 2 & -0.081 & 0.620 & 8 & -0.108 & -0.902 & 1.878 & -1.093 & 0.228 \\
- & 0.9 & 0.6 & 3 & -0.088 & 0.786 & 8 & -0.851 & 0.177 & 0.670 & -0.405 & 0.068 \\
- & 0.9 & 0.6 & 4 & -0.096 & 0.824 & 8 & -1.049 & 0.573 & 0.211 & -0.182 & 0.030 \\
- & 0.9 & 0.6 & 5 & -0.102 & 0.823 & 8 & -1.133 & 0.766 & -0.019 & -0.078 & 0.014 \\
- & 0.9 & 0.6 & 10 & -0.110 & 0.851 & 8 & -1.273 & 1.023 & -0.280 & 0.026 & 0.000 \\
- & 0.9 & 0.6 & 20 & -0.118 & 0.797 & 8 & -1.234 & 0.965 & -0.258 & 0.023 & 0.000 \\
- & 0.9 & 0.6 & 40 & -0.125 & 0.748 & 8 & -1.170 & 0.898 & -0.236 & 0.021 & 0.000 \\
- & 1.0 & 0.65 & 2 & -0.104 & 0.830 & 8 & -1.186 & 1.224 & 0.172 & -0.420 & 0.112 \\
- & 1.0 & 0.65 & 3 & -0.104 & 1.000 & 8 & -1.961 & 2.217 & -0.720 & 0.008 & 0.026 \\
- & 1.0 & 0.65 & 4 & -0.104 & 1.060 & 8 & -2.232 & 2.584 & -1.036 & 0.148 & 0.000 \\
- & 1.0 & 0.65 & 5 & -0.104 & 1.072 & 8 & -2.385 & 2.718 & -1.088 & 0.153 & 0.000 \\
- & 1.0 & 0.65 & 10 & -0.104 & 1.057 & 8 & -2.594 & 2.863 & -1.120 & 0.150 & 0.000 \\
- & 1.0 & 0.65 & 20 & -0.104 & 1.037 & 8 & -2.646 & 2.882 & -1.111 & 0.146 & 0.000 \\
- & 1.0 & 0.65 & 40 & -0.104 & 1.026 & 8 & -2.658 & 2.880 & -1.105 & 0.144 & 0.000 
\end{tabular} 
\caption{This table lists the parameters used in the fitting function to generate the results shown in the figures within the main text. The parameters are identified by the figure in which the binding potential is used, the title of that curve in the figure and some additional identification where applicable. Values are rounded to three decimal places and the value of $a_1$ is enforced.}
\end{table*}

\section{Particle interaction potentials}\label{app:appB}
The net interaction potential between a fluid particle and a planar wall made of particles interacting with the fluid particle via a Lennard-Jones-like potential, with the same form as in Eq.~\eqref{eq:pair_pot} is
\begin{equation}\label{eq:external}
V_\boldi = \epsilon_{wf} \sum_{i'=-\infty}^0 \sum_{j'=-\infty}^\infty \sum_{k'=-\infty}^\infty ((i-i')^2 + j'^2 + k'^2)^{-3},
\end{equation}
where we assume the surface of the wall is located in the plane $i=0$. The parameter $\epsilon_{wf}$ characterises the strength of the interaction between a single wall particle and a fluid particle. The triple sum above spans all of the particles in the wall; i.e.\ the wall is modelled as being discretised on a lattice just as the fluid is and that a wall particle is present on every lattice site where $i \leq 0$. The sums in Eq.\ \eqref{eq:external} can be simplified to obtain
\begin{equation}\label{eq:netInteraction}
V_\boldi = 
\begin{cases}
-\epsilon_w/i^3 & \mbox{ for } i \geq 1 \\
\infty & \mbox{ for } i < 1
\end{cases}
\end{equation}
where the parameter $\epsilon_w$ determines the net strength of attraction to the wall.

The (2D) effective fluid-fluid particle interactions are governed in a similar manner by the potential 
\begin{equation}\label{eq:weights}
\epsilon_{\boldi,\boldj} = \epsilon \left[ (i'^2 + j'^2)^{-3} + 2 \sum_{k'=1}^\infty (i'^2+j'^2+k'^2)^{-3} \right].
\end{equation}
The parameter $\epsilon$ is the strength of a single Lennard-Jones pair interaction between two fluid particles and $\epsilon_{\boldi,\boldj}$ is the weighted interaction between two particles taking into account all of the interactions in the invariant $k$ dimension. In the right hand side of Eq.\,\eqref{eq:weights}, $i'$, $j'$, and $k'$ are the distances between a pair of lattice sites in the $i$, $j$ and $k$ directions respectively.

Due to the fact that calculating long ranged particle interactions can be time consuming, making computer calculations rather slow, it is often necessary to truncate the interactions to some interaction range $L \sigma$. When particle interactions are truncated at a range of $L \sigma$ the external potential Eq.\,\eqref{eq:external} becomes
\begin{widetext}
\begin{equation}\label{eq:truncExternal}
V_\boldi = \epsilon_{wf} \sum_{i'=-L}^0 \sum_{j'=-L}^L \sum_{k'=-L}^L
\begin{cases}
 ((i-i')^2 + j'^2 + k'^2)^{-3} & \text{ for }  ((i-i')^2 + j'^2 + k'^2)\leq L^2 \\
0 & \text{ otherwise}
\end{cases}
\end{equation}
and the fluid-fluid particle interaction weights \eqref{eq:weights} are then given as
\begin{equation}\label{eq:truncWeights}
\epsilon_{\boldi,\boldj} = 
\begin{cases}
 \epsilon \left[ (i'^2 + j'^2)^{-3} + 2 \sum\limits_{k'=1}^{k' \leq \sqrt{L^2-i'^2-j'^2}} (i'^2+j'^2+k'^2)^{-3} \right] & \text{ for } |{\bf i}-{\bf j}|\leq L \\
 0 & \text{ otherwise. }
\end{cases}
\end{equation}
\end{widetext}

\end{appendix}



\end{document}